\documentclass[journal,onecolumn]{IEEEtran} 

\usepackage{amssymb}
\usepackage{amsxtra}
\usepackage{amsmath}
\usepackage{amsthm}
\usepackage{mathtools}
\usepackage{xcolor}

\usepackage{tikz}
\usetikzlibrary{arrows}
\usetikzlibrary{shapes}
\usepackage{pgfplots}
\pgfplotsset{compat=newest} 
\pgfplotsset{plot coordinates/math parser=false}
\usetikzlibrary{plotmarks}
\usetikzlibrary{spy}

\allowdisplaybreaks

\newtheorem{theorem}{Theorem}
\newtheorem{cor}[theorem]{Corollary}

\def\cC{\mathcal{C}}
\def\cF{\mathcal{F}}
\def\cX{\mathcal{X}}
\def\cXN{\mathcal{X}_{d,k}^N}
\def\cY{\mathcal{Y}}
\def\cQ{\mathcal{Q}}
\def\e{\epsilon}
\def\a{\alpha}
\def\b{\beta}
\def\l{\lambda}
\def\pr{\text{Pr}}
\def\op{\overline{p}}
\def\ova{\overline{a}}
\def\ovb{\overline{b}}
\def\|{\,||\,}
\def\bl{\bigl(}
\def\br{\bigr)}
\def\ob{\overline{\beta}}
\def\oa{\overline{\alpha}}
\def\pd{\partial}

\begin{document}
\title{Dual Capacity Upper Bounds for Noisy Runlength Constrained Channels}
\author{Andrew Thangaraj\thanks{A. Thangaraj is with the Department of
    Electrical Engineering, Indian Institute of Technology Madras,
    Chennai 600036, India, Email: andrew@ee.iitm.ac.in. Parts of this paper will appear in the IEEE Information Theory
Workshop 2017, Cambridge, UK, Sep 11-14, 2017.}}
\maketitle
\begin{abstract}
Binary-input memoryless channels with a runlength constrained input
are considered. Upper bounds to the capacity of such noisy runlength
constrained channels are derived using the dual capacity
method with Markov test distributions satisfying the
Karush-Kuhn-Tucker (KKT) conditions for the capacity-achieving output distribution. Simplified algebraic characterizations of the bounds are presented for the binary
erasure channel (BEC) and the binary symmetric channel (BSC). These
upper bounds are very close to achievable rates, and improve upon
previously known feedback-based bounds for a large range of
channel parameters. For the binary-input Additive White Gaussian Noise
(AWGN) channel, the upper bound is simplified to a small-scale
numerical optimization problem. These results provide some of the
simplest upper bounds for an open capacity problem that has theoretical and
practical relevance.
\end{abstract}
\section{Introduction}
Runlength constrained input is commonly used in data storage
applications, and channels with runlength constraints have been widely
studied in information theory \cite{handbook}. Characterizing the capacity of noisy channels with runlength-constrained input has been an open problem for quite some time
now. Simulation-based methods for approximately computing lower and
upper bounds to the noisy constrained capacity are
well-known \cite{1661831}\cite{955166}. Numerical methods have been
proposed for computing the capacity in some cases \cite{6875399}. More recently, the feedback
capacity, which serves as an upper bound to the non-feedback case, has
been characterized as a computable optimization problem
\cite{7308065}. See references in \cite{7308065} for a more complete bibliography of this area.

In this work, we derive upper bounds for noisy channels
with runlength-constrained input. The main idea is the use of the dual
capacity upper bound \cite{topsoe67}\cite{KEMPERMAN1974101}\cite{Csiszar11} and tuning
it to the scenario of input-constrained noisy channels. The dual
capacity bound has been used by several authors in applications such
as optical channels \cite{LMW2009}, MIMO channels \cite{1237131},
phase noise channels \cite{6218668} and peakpower-limited Gaussian
channels \cite{7282870}. In \cite{955166}, the dual bound was used for
runlength-constrained channels. 

In the dual bound, a critical choice is that of the test distribution
on the output alphabet. A Markov test distribution on the output of a runlength-constrained
channel was used in \cite{955166}. The main innovation in this work is enforcing the Karush-Kuhn-Tucker (KKT) conditions for the
capacity-achieving output distribution on the test distribution. This
is done by equating suitably defined metrics of cycles in the state
diagram of the constraint. As
shown, this results in tight bounds and interesting and simple algebraic
characterizations of the upper bound in several examples such as the
runlength-constrained binary erasure channel and binary symmetric channel.
For the binary-input Additive White Gaussian Noise (AWGN) channel, the
bound can be computed by a small-scale optimization problem. 

When compared to the feedback-based bound in \cite{7308065}, the dual
capacity method in this paper is more direct and easily applicable to
general $(d,k)$ constraints and channels with continuous output. The
resulting bounds are tighter and simpler algebraic characterizations in many cases. 
When compared to \cite{955166}, the use of KKT conditions in test
distributions is novel.

The rest of this paper is organized as follows. Section \ref{sec:preliminaries}
 introduces the notation and problem setup. Section
 \ref{sec:upper-bounds} provides the main results. The proofs and
 computations of bounds are shown in Section
 \ref{sec:proofs-computations}, and concluding remarks are made in
 Section \ref{sec:concluding-remarks}.
\section{Notation and Definitions}
\label{sec:preliminaries}
For integers $a$, $b$, the notation $[a:b]$ denotes the set of integers
$\{a,a+1,\ldots,b\}$. The set $[a:b]$ is empty if $a>b$. Given a
sequence $v=(v_1,v_2,\ldots)$, and positive integers $a$, $b$ the notation $v_{[a:b]}$ denotes the sub-sequence $(v_i:i\in[a:b])$. The sequence $v_{[1:N]}$ is also denoted by $v^N$. For a fraction $x$, we denote
$\overline{x}=1-x$. 

A directed graph $G=(V,E)$ consists of a set of vertices $V$ and a set
of directed edges $E\subseteq V\times V$. An edge $e=(v_1,v_2)$ is
directed from $v_1$ to $v_2$. We
will use the terms edges or arcs interchangeably for the directed edges of a directed graph. We will consider directed
graphs that may have self-loops, i.e., edges of the form $(v,v)$, but
we will not consider graphs with multiple parallel edges between the
same pair of vertices in the same direction. A walk of length $k$ in a directed
graph $G=(V,E)$ is an alternating sequence of vertices and edges
$(v_1,e_1,v_2,e_2,v_3,\ldots,e_k,v_{k+1})$ such that
$e_i=(v_i,v_{i+1})\in E$. Often a walk is denoted simply by the
sequence of vertices $(v_1,v_2,\ldots)$ or the sequences of edges
$(e_1,e_2,\ldots)$. A walk with distinct vertices is called a path. A walk of
length $k$ with distinct $v_1$, $v_2$, $\ldots$, $v_k$ and
$v_1=v_{k+1}$ is called a cycle of length $k$.

\subsection{Runlength-constrained channels}
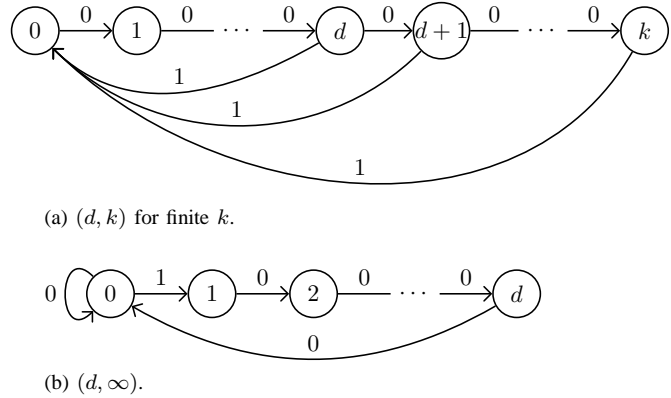
\begin{figure}[htb]
  \centering
    \tikzstyle{cnode}=[circle,draw]
  \begin{tikzpicture}
    \begin{scope}[scale=0.7, every node/.style={scale=0.9},node
      distance=1.5cm,>=angle 90,semithick]
    \node[minimum size=0.7cm,cnode,inner sep=0] (n0)  {$0$};
    \node[minimum size=0.7cm,cnode,inner sep=0] (n1) [right of=n0]     {$1$};
    \node (ndots1) [right of=n1]     {$\cdots$};
    \node[minimum size=0.7cm,cnode,inner sep=0] (nd) [right of=ndots1]     {$d$};
    \node[minimum size=0.7cm,cnode,inner sep=0] (ndp1) [right of=nd]     {$d+1$};
    \node (ndots2) [right of=ndp1]     {$\cdots$};
    \node[minimum size=0.7cm,cnode,inner sep=0] (nk) [right of=ndots2]     {$k$};
    \draw[->] (n0) --node[above]{$0$} (n1);
    \draw[-] (n1) --node[above]{$0$} (ndots1);
    \draw[->] (ndots1) --node[above]{$0$} (nd);
    \draw[->] (nd) --node[above]{$0$} (ndp1);
    \draw[-] (ndp1) --node[above]{$0$} (ndots2);
    \draw[->] (ndots2) --node[above]{$0$} (nk);
    \draw[->] (nd)     to[out=210,in=315] node[above]{$1$} (n0);
    \draw[->] (ndp1) to[out=225,in=315] node[above]{$1$} (n0);
    \draw[->] (nk)     to[out=240,in=315] node[above]{$1$} (n0);
    \end{scope}
    \node[anchor=west] at (0,-2.5) {\footnotesize (a) $(d,k)$ for finite $k$.};
    \begin{scope}[shift={(1,-3.5)},scale=0.7, every node/.style={scale=0.9},node distance=1.5cm,>=angle 90,semithick]
    \node[minimum size=0.7cm,cnode] (n3)  {$0$};
    \node[minimum size=0.7cm,cnode] (n4) [right of=n3]     {$1$};
    \node[minimum size=0.7cm,cnode] (n5) [right of=n4]     {$2$};
    \node (n6) [right of=n5]     {$\cdots$};
    \node[minimum size=0.7cm,cnode] (n7) [right of=n6]     {$d$};
    \draw[->] (n3) edge[loop left,out=135,in=225,min distance=1cm] node[left](l1){$0$} (n3);
    \draw[->] (n3) --node[above]{$1$} (n4);
    \draw[->] (n4) --node[above]{$0$} (n5);
    \draw[-] (n5) --node[above]{$0$} (n6);
    \draw[->] (n6) --node[above]{$0$} (n7);
    \draw[->] (n7) to[out=210,in=330] node[above]{$0$} (n3);
   \end{scope}
    \node[anchor=west] at (0,-4.7) {\footnotesize (b) $(d,\infty)$.};
  \end{tikzpicture}  
  \caption{Directed graph $G_{d,k}$: State diagram for $(d,k)$-constrained sequences.}
\label{fig:dk}
\end{figure}

For non-negative integers $d$ and $k$ with $k>d$, a $(d,k)$-constrained binary sequence is a
sequence of bits for which there are at least $d$ zeros and at most $k$
zeros ($k$ can be infinity) between any two 1s. A $(d,k)$-constrained sequence is usually
represented as a walk on a state diagram. The state diagram for
finite $k$ and infinte $k$ are shown in Fig. \ref{fig:dk}.
As seen, the state diagram for $(d,k)$-constrained sequences is a
directed graph with edge labels, which we will denote $G_{d,k}$. 

Let $\cXN$ denote the set of all $(d,k)$-constrained binary sequences
of length $N$. The capacity of $(d,k)$-constrained sequences, denoted
$C_{d,k}$, is defined and characterized as follows:
\begin{gather}
  C_{d,k}\triangleq \lim_{N\to\infty}\dfrac{\log_2|\cXN|}{N} = \log_2(1/\lambda),\label{eq:33}\\
\lambda\in(0,1) \text{ solves } z^{k+2}-z^{d+1}-z+1=0.\nonumber
\end{gather}
For $k=\infty$, the above characterization holds with the term
$z^{k+2}$ set as $0$.

Consider a binary-input memoryless channel with input alphabet
$\cX=\{0,1\}$, output alphabet $\cY$ and transition probability
denoted $p_{Y|X}(y|x)$. All channels considered in this
paper will have binary input. The output alphabet $\cY$ may be either
discrete or continuous. For discrete $\cY$, $p_{Y|X}$ will be denoted
as an $|\cX|\times|\cY|$ matrix with the $(x,y)$-th entry being $p_{Y|X}(y|x)$. For continuous $\cY$, we will specify the conditional
probability density function (PDF) $p_{Y|X=x}$ for $x=0,1$. The capacity of a channel $p_{Y|X}$, denoted $C(p_{Y|X})$, is given by $C(p_{Y|X})=\max_{p_X}I(X;Y)$. 
Standard convex optimization methods can be used for computing the
capacity. Explicit algebraic expressions or small-scale numerical
computations are available for computing the capacity of
standard channels such as the binary erasure channel (BEC), binary symmetric
channel (BSC) and the binary-input additive white Gaussian noise
(BIAWGN) channel.

By a $(d,k)$-constrained channel $p_{Y|X}$, we refer to a channel
$p_{Y|X}$ whose input is constrained to be a $(d,k)$-constrained
binary sequence. Specifically, if the channel $p_{Y|X}$ is used $N$
times, its input $X^N=[X_1, X_2, \ldots,X_N]$ is a length-$N$,
$(d,k)$-constrained binary sequence, i.e. $X^N\in\cXN$. The output $Y^N=[Y_1, Y_2, \ldots, Y_N]$ obeys
a memoryless channel transition law
$p(y^N|x^N)=\prod_{i=1}^Np_{Y|X}(y_i|x_i)$.
The capacity of the $N$-letter, $(d,k)$-constrained channel $p_{Y|X}$,
denoted $C^N_{d,k}(p_{Y|X})$, is given by
\begin{equation}
  \label{eq:1}
  C^N_{d,k}(p_{Y|X})=\max_{p(x^N): x^N\in \cX_{d,k}^N}I(X^N;Y^N).
\end{equation}
The capacity of the $(d,k)$-constrained channel
$p_{Y|X}$, denoted $C_{d,k}(p_{Y|X})$, is defined
as
\begin{equation}
  C_{d,k}(p_{Y|X})=\lim_{N\to\infty}\frac{1}{N}C^N_{d,k}(p_{Y|X}).
\label{eq:18}
\end{equation}
Characterizing the $(d,k)$-constrained capacity of channels has proven
to be considerably more difficult because the memory in the input
makes the computation of \eqref{eq:1} dependent on $N$, which grows to
infinity. The main result of this paper is the derivation of upper bounds
for $C_{d,k}(p_{Y|X})$ that either have simple algebraic
characterizations (like that of $C_{d,k}$ in \eqref{eq:33})  or
small-scale numerical computation procedures. In particular, the
computations are independent of $N$. The bounds, in many cases, are
seen to be extremely close to achievable rates showing that they are tight.

In the rest of the paper, we will refer to the channel $p_{Y|X}$ with $(d,k)$-constrained input
simply as the $(d,k)$-constrained $p_{Y|X}$. Standard channels
considered are the following:
\begin{enumerate}
\item BEC$(\epsilon)$ with
$$p_{Y|X}=\begin{bmatrix}1-\e&\e&0\\0&\e&1-\e\end{bmatrix}.$$
\item BSC$(p)$ with
  $$p_{Y|X}=\begin{bmatrix}1-p&p\\p&1-p\end{bmatrix}.$$
\item BIAWGN$(\sigma^2)$ with 
$$p_{Y|X=x}\sim N((-1)^x,\sigma^2),$$
 where $N(m,s)$ denotes the Gaussian distribution with
mean $m$ and variance $s$. 
\end{enumerate}
We will use the notation $(d,k)$-BEC$(\e)$, $(d,k)$-BSC$(p)$ and $(d,k)$-BIAWGN$(\sigma^2)$, respectively, for the
$(d,k)$-constrained versions, and the notation $C_{d,k}(\epsilon)$,
$C_{d,k}(p)$ and $C_{d,k}(\sigma)$ for the $(d,k)$-constrained
capacities of the standard channels.

\subsection{State diagram with memory $\mu$ for the $(d,k)$-constraint}
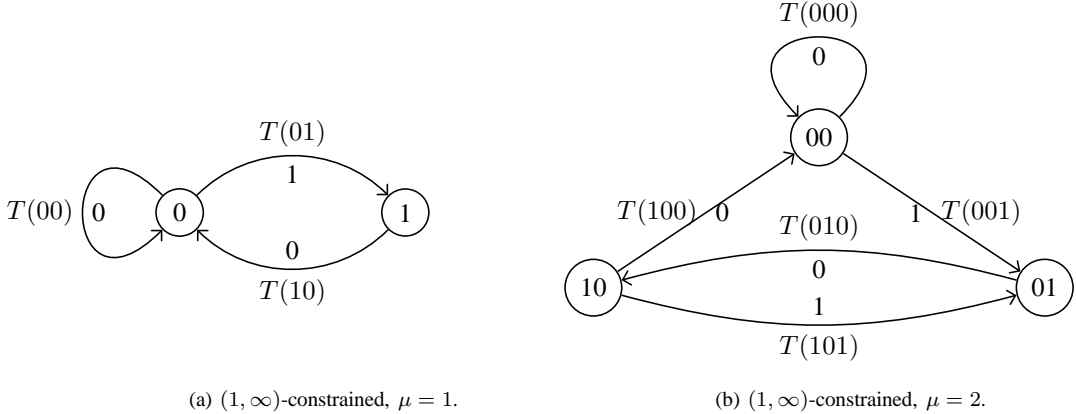
\begin{figure*}[htb]
  \centering
  \tikzstyle{cnode}=[circle,draw]
  \begin{tikzpicture}
    \begin{scope}[node distance=2cm,>=angle 90,semithick]
    \node[cnode] (n1)                   {0};
    \node[cnode] (n2) [right of=n1,xshift=1cm]     {1};
    \draw[->] (n1) edge[loop left,out=135,in=225,min distance=2cm]
    node[left]{$T(00)$} node[right]{0} (n1);
    \draw[->] (n1) to[out=45,in=135] node[above]{$T(01)$} node[below]{1} (n2);
    \draw[->] (n2) to[out=225,in=315] node[below]{$T(10)$} node[above]{0} (n1);
    \end{scope}
    \node[anchor=west] at (0,-2.5) {\footnotesize (a) $(1,\infty)$-constrained, $\mu=1$.};
        \begin{scope}[shift={(8.5,1)},node distance=3cm,>=angle 90,semithick]
    \node[cnode] (n00)                   {00};
    \node[cnode] (n01) [right of=n00,yshift=-2cm]     {01};
    \node[cnode] (n10) [left of=n00,yshift=-2cm]     {10};
    \draw[->] (n00) edge[loop left,out=45,in=135,min distance=2cm]
    node[above]{$T(000)$} node[below]{0} (n00);
    \draw[->] (n00) -- node[right]{$T(001)$} node[left]{1} (n01);
    \draw[->] (n01) to[out=165,in=15] node[above]{$T(010)$}
    node[below]{0} (n10);
    \draw[->] (n10) to[out=345,in=195] node[below]{$T(101)$}
    node[above]{1} (n01);
    \draw[->] (n10) -- node[left]{$T(100)$} node[right]{0} (n00);
    \end{scope}
    \node[anchor=west] at (7,-2.5) {\footnotesize (b) $(1,\infty)$-constrained, $\mu=2$.};
  \end{tikzpicture} 
\caption{$G^{\mu}_{1,\infty}$: Memory-$\mu$ state diagram for the $(1,\infty)$ constraint.}
\label{fig:sdmem}
\end{figure*}

Let $\mu$ be a positive integer satisfying
\begin{equation}
  \mu\ge\begin{cases}
k,&k:\text{ finite},\\
d,&k=\infty.
\end{cases}
\end{equation}
The state diagram for the $(d,k)$ constraint with memory $\mu$ is an
edge-labeled, directed graph, denoted $G^{\mu}_{d,k}$, and defined as follows. The
vertex set of $G^{\mu}_{d,k}$ is the set $\cX^{\mu}_{d,k}$ of $(d,k)$-constrained sequences
of length $\mu$. From a vertex $(x_1x_2\ldots x_{\mu})$, an arc with label $x_{\mu+1}\in\{0,1\}$ is drawn
whenever $(x_1x_2\ldots x_{\mu}x_{\mu+1})$ is a valid length-$(\mu+1)$, $(d,k)$-constrained
sequence. The arc ends in the vertex
$(x_2\ldots x_{\mu+1})$. Examples of state diagrams with memory are shown in Fig. \ref{fig:sdmem}.

In $G^{\mu}_{d,k}$, the length-$l$ cycle $\{v_1,v_2,\ldots,v_l,v_1\}$ is considered \emph{equivalent} to the
cycle $\{v_i,v_{i+1},\ldots,v_l,v_1,\ldots,v_{i-1},v_i\}$, which is
simply the same cycle traversed with a different starting point. Let
the set of cycles of $G^{\mu}_{d,k}$ be denoted $\cC^{\mu}_{d,k}$ with
the convention that no two cycles in it are equivalent. The length of
a cycle $c\in\cC^{\mu}_{d,k}$ is denoted $l(c)$.

Consider a length-$l$ cycle $c=\{v_1,v_2,\ldots,v_l,v_1\}\in
\cC^{\mu}_{d,k}$ with $e_i=(v_i,v_{i+1})$, $1\le i\le l$. Suppose
$v_1=(x_1x_2\ldots x_{\mu})\in\cX^{\mu}_{d,k}$ and let the label of
edge $e_i$ be denoted $x_{\mu+i}$. The sequence
$(x_1\ldots x_{\mu}x_{\mu+1}\ldots x_{\mu+l})$ is a valid
$(d,k)$-sequence of length $\mu+l$, which we will denote $x(c)$ and associate with the cycle $c$. For
example, in $G^{2}_{1,\infty}$ shown in Fig. \ref{fig:sdmem}, the length-3
cycle $c=\{00,01,10,00\}$ is associated with the length-5 $(1,\infty)$-sequence $x(c)=(00100)$.

\subsection{Markov test distributions}
The upper bound on $C_{d,k}(p_{Y|X})$ is expressed using certain Markov test distributions on sequences of channel outputs. A sequence or chain of random variables $(Y_1,Y_2,\ldots)$ (with $Y_i$ taking values in $\cY$) is said to be Markov with memory $\mu$ if
$Y_i$ is conditionally independent of $Y_{[1:i-\mu-1]}$ given
$Y_{[i-\mu:i-1]}$. The distribution of a Markov chain is specified by
the transition probability distributions
$\pr(Y_{\mu+1}=y_{\mu+1}|Y^{\mu}=y^{\mu})$ and the initial
distribution $\pr(Y^{\mu}=y^{\mu})$ for all $y^{\mu+1}\in\cY^{\mu+1}$. The distributions are specified as Probability Mass Functions (PMFs) or Probability
Density Functions (PDFs) depending on whether $\cY$ is discrete or continuous.

Let $\cQ_{\mu}$ refer to the collection of transition probabilities of
Markov chains $(Y_1,Y_2,\ldots)$ with memory $\mu$. A specific $q\in\cQ_{\mu}$ is
specified by providing 
$$q(y_{\mu+1}|y^{\mu})\triangleq \pr(Y_{\mu+1}=y_{\mu+1}|Y^{\mu}=y^{\mu})$$
for $y^{\mu+1}\in \cY^{\mu+1}$.
For a Markov chain $(Y_1,Y_2,\ldots)$ with transition probability $q$
and a channel output sequence $y^N\in\cY^N$, we have
\begin{equation}
  \pr(Y^N=y^N)=\prod_{i=1}^{\mu}q(y_i|y^{i-1})\prod_{i=\mu+1}^Nq(y_i|y_{[i-\mu:i-1]}),
\end{equation}
where, for $1\le i\le \mu$,
$$q(y_i|y^{i-1})\triangleq\pr(Y_i=y_i|Y^{i-1}=y^{i-1})$$
is specified by the initial distribution. We will refer to the set of
transition probabilities $q$ as Markov distributions of memory $\mu$
on the alphabet $\cY$.

\subsection{Metric in the state diagram}
Let $D\bl p_1(\cdot)\|
p_2(\cdot)\br=\sum_{y\in\cY}p_1(y)\log_2\dfrac{p_1(y)}{p_2(y)}$ denote
the relative entropy of two distributions $p_1$ and $p_2$ on the same
alphabet $\cY$. Given a Markov distribution $q$ of memory $\mu$ on the
channel output alphabet $\cY$, we associate a metric to every edge in the
graph $G^{\mu}_{d,k}$. The metric for the edge from vertex $(x_1\ldots
x_{\mu})$ with label
$x_{\mu+1}$, denoted $T_{q,p_{Y|X}}(x_1\ldots x_{\mu+1})$, is
defined as follows:
\begin{align}
T_{q,p_{Y|X}}(x_1\ldots x_{\mu+1})&=\sum_{y^{\mu}\in\cY^{\mu}}p_{Y^{\mu}|X^{\mu}}(y^{\mu}|x^{\mu})D\bl p_{Y|X}(\,\cdot\,|x_{\mu+1})\| q(\,\cdot\,|y^{\mu})\br.\label{eq:10}
\end{align}
In Fig. \ref{fig:sdmem}, the edge metrics are shown on the
edges of the state diagram. The subscripts $q$ and $p_{Y|X}$ are dropped in the
notation for the edge metric when it is either
clear from the context or not important. 
\subsubsection{Example - BEC$(\e)$}
To further illustrate, let us
consider the channel BEC$(\e)$, memory $\mu=1$, and the Markov distribution
\begin{equation}
   q(y_2|y_1)=\begin{bmatrix}
\b(1-\e)&\e&\ob(1-\e)\\
\a(1-\e)&\e&\oa(1-\e)\\
1-\e&\e&0
\end{bmatrix},
   \label{eq:4a}
\end{equation}
where the rows correspond to $y_1=0,?,1$, the columns correspond to $y_2=0,?,1$
in that order ($?$ denotes erasure symbol), and $\a,\b$ are parameters satisfying
$0\le\a,\b\le1$ and $\ob=1-\b$, $\oa=1-\a$. For this example, we see that
\begin{align}
T(00)&=\sum_{y_1\in\{0,?,1\}}p_{Y|X}(y_1|0)D\bl \,p_{Y|X}(\,\cdot\,|0)\| q(\,\cdot\,|y_1)\,\br\nonumber\\
&=(1-\e)\, D\bl\, [1-\e, \e, 0]\| [\b(1-\e), \e, \ob(1-\e)]\,\br\nonumber\\
&\quad +\ \, \e\quad\, D\bl\, [1-\e, \e, 0]\| [\a(1-\e), \e, \oa(1-\e)]\,\br\nonumber\\
&=(1-\e)^2\log_2(1/\b)+\e(1-\e)\log_2(1/\a).\label{eq:26}
\end{align}
Similar expressions derived for $T(01)$ and $T(10)$ are
\begin{align}
  T(01)&=(1-\e)^2\log_2(1/\ob)+\e(1-\e)\log_2(1/\oa),\label{eq:27}\\
  T(10)&=\e(1-\e)\log_2(1/\a).\label{eq:28}
\end{align}
\subsubsection{Metric of walks and cycles}
A valid $(d,k)$ sequence $x^N\in\cX^N_{d,k}$ corresponds to the walk
in $G^{\mu}_{d,k}$ of length $N-\mu$ with sequence of vertices $(x_{[1:\mu]},x_{[2:\mu+1]},\ldots,x_{[N-\mu+1:N]})$. The metric of a walk in $G^{\mu}_{d,k}$ is defined to be the sum of the
metrics on the edges in the walk. Since a cycle is a walk, the same
definition extends for cycles as well. Consider a cycle $c$ of length $l(c)$
in $G^{\mu}_{d,k}$ associated to the sequence
$(x_1\ldots x_{\mu+l(c)})$. The metric of the cycle, denoted
$T_{q,p_{Y|X}}(c)$, is readily seen to be
\begin{equation}
  T_{q,p_{Y|X}}(c)=\sum_{i=1}^{l(c)}T_{q,p_{Y|X}}(x_i\ldots x_{\mu+i}).
\end{equation}
As an example, for the cycle in
$G^2_{1,\infty}$ associated to the sequence $(00100)$, the metric is $T(001)+T(010)+T(100)$.
\section{Upper bounds}
\label{sec:upper-bounds}
The main upper bound and several examples are presented and discussed in this section with the
proofs and details of computations to be provided later. A dual capacity upper bound for the
capacity of the $(d,k)$-constrained
channel $p_{Y|X}$ is given in the following theorem and its corollary, which use the
notation described in Section \ref{sec:preliminaries}.
\begin{theorem}[Dual bound]
Let $q$ be a Markov distribution with memory $\mu$ over the output alphabet
$\cY$ satisfying the constraint that 
\begin{equation}
\dfrac{T_{q,p_{Y|X}}(c)}{l(c)}=\dfrac{T_{q,p_{Y|X}}(c')}{l(c')}  
\label{eq:9}
\end{equation}
for any two cycles $c,c'\in\cC^{\mu}_{d,k}$. Then, the capacity of the
$(d,k)$-constrained channel $p_{Y|X}$ is upper bounded by the common
value in \eqref{eq:9}, i.e., 
$$C_{d,k}(p_{Y|X})\le \dfrac{T_{q,p_{Y|X}}(c)}{l(c)}$$
for any $c\in\cC^{\mu}_{d,k}$.
\label{thm:ubq}
\end{theorem}
Since the upper bound holds for every $q$ satisfying the constraint
\eqref{eq:9}, the bound can be improved by minimizing over a set of
Markov distributions satisfying the constraint. This observation
results in the following corollary, which is useful in deriving bounds
for standard channels.
\begin{cor}[Dual bound]
Let $Q$ be a collection of Markov distributions with memory $\mu$ over the output alphabet
$\cY$. Then,
\begin{align}
&C_{d,k}(p_{Y|X})\le \min_{q\in Q}\ t(q,p_{Y|X})\\
&\text{subject to }t(q,p_{Y|X})=\dfrac{T_{q,p_{Y|X}}(c)}{l(c)},\quad\forall c\in\cC^{\mu}_{d,k}.\nonumber
\end{align}
\label{thm:ubQ}
\end{cor}
The main observation about the dual upper bounds in Theorem \ref{thm:ubq}
and Corollary \ref{thm:ubQ} is that they are independent of $N$, the
length of the channel input, occurring in the definition of
$C_{d,k}(p_{Y|X})$. In the rest of this paper, we will refer to the above
theorem and corollary as the dual bound theorem and the dual bound
corollary, respectively. We will show by several examples that the
dual bound theorem and corollary result in simple algebraic
characterizations or small scale numerical computations for the
upper bound.  
\subsection{Binary erasure channel}
For the $(d,k)$-BEC$(\e)$ with specific values of $d$ and $k$, upper
bounds on the capacity are given in the following theorems. Recall that
the capacity is denoted $C_{d,k}(\e)$.

\begin{theorem}
(1) For the $(1,\infty)$-BEC$(\e)$, the following upper bound holds:
  \begin{align}
            C_{1,\infty}(\e)\le &(1-\e)^2\log_2(1/\b^*)+\e(1-\e)\log_2(2-\b^*),\\
&\b^*\in(0,1]\text{ solves }\b^{2(1-\e)}=1-\b.\nonumber
           \end{align}
(2) For the $(1,\infty)$-BEC$(\e)$, the following upper bound holds:
  \begin{align}
&  C_{(1,\infty)}(\e)\le (1-\e)\left[(1-\e)^2\log_2\frac{1}{\b}+\e(1-\e)\log_2\frac{\a^2}{\gamma_1(2\a-1)}
  +\e^2\log_2\frac{1}{\a}\right],
\end{align}
where $\gamma_1=1-\a-\b+2\a\b$, and $\alpha,\beta\in[0,1]$ solve
\begin{align}
&(1-\e)^2\log_2\frac{\b^2(1-\a)}{\gamma_2}+\e(1-\e)\log_2\frac{(2\a-1)^2\gamma_1}{\a^2\gamma_2}\nonumber\\
&=(1-\e)^2\log_2\frac{\b^2(1-\a)}{(1-\b)^2(2\a-1)}+2\e(1-\e)\log_2\frac{\gamma_1}{\a(1-\b)}\nonumber\\
&=\e^2\log_2\frac{1-\alpha}{\alpha}
\end{align}
with $\gamma_2=2-3\a-\b+2\a\b$.
\label{thm:bec}
\end{theorem}
The first part of the above theorem is obtained by using the dual
bound corollary with $\mu=1$ and the collection of test distributions in
\eqref{eq:4a}. Note that the expression is reminiscent of the
noiseless capacity of $(1,\infty)$ sequences, which is given by
\begin{align}
  C_{1,\infty}&=\log_2(1/\b),\\
\b=(\sqrt{5}-1)/2& \text{ solves } \b^2=1-\b.\nonumber
\end{align}
The second part uses a collection of memory-2 Markov test
distributions and involves more computations. The details are provided
in later sections.

\begin{figure}[htb!]
  \centering
  \input{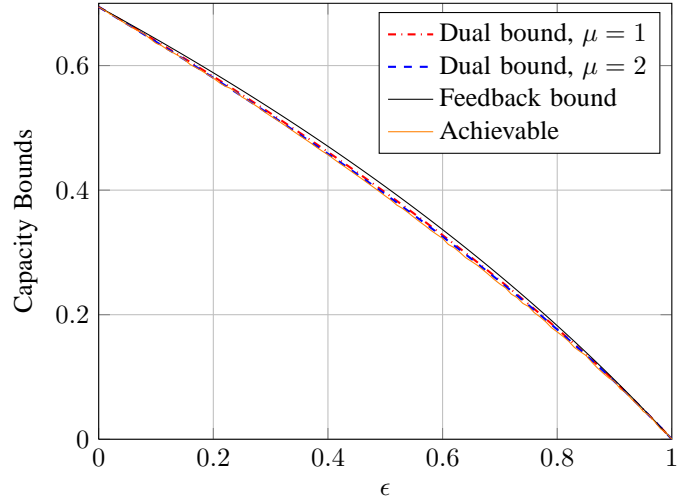}
  \caption{Capacity upper bounds for $(1,\infty)$-constrained BEC$(\epsilon)$.}
  \label{fig:bec}
\end{figure}

Fig. \ref{fig:bec} shows plots of the dual bounds from Theorem
\ref{thm:bec}. 
For comparison, a feedback-based upper bound from
\cite{7308065} and an achievable rate computed using the simulation
method of \cite{1661831} are shown. While all four lines are
reasonably close in the figure, the zoomed inset shows that the dual
capacity bounds are better with the $\mu=2$ bound almost meeting the
achievable rate.

\begin{figure}[htb!]
  \centering
%
%
%
\definecolor{mycolor1}{rgb}{0.00000,0.44700,0.74100}%
\definecolor{mycolor2}{rgb}{0.85000,0.32500,0.09800}%
\definecolor{mycolor3}{rgb}{0.92900,0.69400,0.12500}%
\begin{tikzpicture}

\begin{axis}[%
width=3in,
height=2.28372652864709in,
scale only axis,
xmin=0,
xmax=1,
xlabel={$\epsilon$},
xmajorgrids,
ymin=0,
ymax=0.45,
ylabel={Capacity Bounds},
ymajorgrids,
legend style={draw=black,fill=white,legend cell align=left,legend pos=south west}
]
\addplot [color=red,dashdotted,thick]
  table[row sep=crcr]{%
0	0.405685231375825\\
0.01	0.40790622186012\\
0.02	0.410026354127841\\
0.03	0.412042491373961\\
0.04	0.413951675795989\\
0.05	0.415751120179121\\
0.06	0.417438200017487\\
0.07	0.419010446128575\\
0.08	0.420465537722056\\
0.09	0.421801295887859\\
0.1	0.423015677471625\\
0.11	0.424106769308546\\
0.12	0.425072782789227\\
0.13	0.425912048733501\\
0.14	0.426623012550246\\
0.15	0.427204229663096\\
0.16	0.427654361183678\\
0.17	0.427972169815481\\
0.18	0.428156515972865\\
0.19	0.428206354100986\\
0.2	0.42812072918349\\
0.21	0.427898773425924\\
0.22	0.427539703103666\\
0.23	0.427042815564094\\
0.24	0.426407486373459\\
0.25	0.425633166599601\\
0.26	0.424719380222363\\
0.27	0.423665721664072\\
0.28	0.422471853433047\\
0.29	0.421137503873557\\
0.3	0.419662465016138\\
0.31	0.418046590522547\\
0.32	0.416289793720061\\
0.33	0.414392045720158\\
0.34	0.412353373616926\\
0.35	0.41017385876088\\
0.36	0.407853635104115\\
0.37	0.40539288761297\\
0.38	0.402791850744642\\
0.39	0.400050806984368\\
0.4	0.397170085440011\\
0.41	0.394150060491064\\
0.42	0.390991150489244\\
0.43	0.387693816508003\\
0.44	0.384258561138438\\
0.45	0.380685927329171\\
0.46	0.376976497267923\\
0.47	0.373130891302576\\
0.48	0.369149766899629\\
0.49	0.365033817638001\\
0.5	0.360783772236238\\
0.51	0.356400393611183\\
0.52	0.35188447796624\\
0.53	0.347236853907369\\
0.54	0.342458381584949\\
0.55	0.337549951859616\\
0.56	0.332512485490186\\
0.57	0.32734693234163\\
0.58	0.322054270611057\\
0.59	0.316635506069435\\
0.6	0.311091671316655\\
0.61	0.305423825047285\\
0.62	0.299633051324013\\
0.63	0.293720458855433\\
0.64	0.287687180274255\\
0.65	0.281534371411409\\
0.66	0.275263210560638\\
0.67	0.268874897727147\\
0.68	0.262370653852462\\
0.69	0.255751720005953\\
0.7	0.249019356531157\\
0.71	0.242174842132161\\
0.72	0.235219472881447\\
0.73	0.228154561125643\\
0.74	0.220981434258997\\
0.75	0.213701433325675\\
0.76	0.206315911400275\\
0.77	0.198826231680195\\
0.78	0.191233765202062\\
0.79	0.183539888065107\\
0.8	0.175745978003734\\
0.81	0.167853410094836\\
0.82	0.159863551305354\\
0.83	0.151777753471321\\
0.84	0.143597344134653\\
0.85	0.135323614422743\\
0.86	0.126957802798298\\
0.87	0.11850107296905\\
0.88	0.109954483424847\\
0.89	0.101318944790543\\
0.9	0.09259515915284\\
0.91	0.0837835322222077\\
0.92	0.0748840436910118\\
0.93	0.0658960516715208\\
0.94	0.0568179900896311\\
0.95	0.047646885740025\\
0.96	0.0383775563884631\\
0.97	0.0290012047747818\\
0.98	0.0195027421530087\\
0.99	0.00985490996197093\\
1	0\\
};
\addlegendentry{Dual bound, $\mu=2$};

\addplot [color=blue,dashed,thick]
  table[row sep=crcr]{%
0	0.405685231375825\\
0.01	0.405683269596883\\
0.02	0.405674952457581\\
0.03	0.405656692224284\\
0.04	0.405624976516721\\
0.05	0.405576363119902\\
0.06	0.405507475425743\\
0.07	0.405414998445289\\
0.08	0.405295675337482\\
0.09	0.405146304405113\\
0.1	0.404963736512825\\
0.11	0.404744872885959\\
0.12	0.404486663252685\\
0.13	0.404186104295116\\
0.14	0.403840238378179\\
0.15	0.403446152527801\\
0.16	0.403000977632524\\
0.17	0.402501887845004\\
0.18	0.401946100162026\\
0.19	0.401330874163592\\
0.2	0.400653511893496\\
0.21	0.399911357865385\\
0.22	0.399101799179864\\
0.23	0.398222265739549\\
0.24	0.397270230550224\\
0.25	0.396243210097412\\
0.26	0.395138764788716\\
0.27	0.393954499453205\\
0.28	0.39268806389003\\
0.29	0.391337153459188\\
0.3	0.389899509708105\\
0.31	0.388372921028342\\
0.32	0.386755223337302\\
0.33	0.385044300780378\\
0.34	0.383238086449441\\
0.35	0.381334563114015\\
0.36	0.379331763961873\\
0.37	0.377227773346163\\
0.38	0.375020727536483\\
0.39	0.372708815471614\\
0.4	0.37029027951189\\
0.41	0.36776341618943\\
0.42	0.365126576954645\\
0.43	0.362378168917653\\
0.44	0.359516655583389\\
0.45	0.356540557579362\\
0.46	0.353448453375133\\
0.47	0.350238979992738\\
0.48	0.346910833707359\\
0.49	0.343462770737665\\
0.5	0.339893607925309\\
0.51	0.336202223403171\\
0.52	0.332387557251962\\
0.53	0.328448612144886\\
0.54	0.324384453980093\\
0.55	0.320194212500686\\
0.56	0.315877081902082\\
0.57	0.311432321426523\\
0.58	0.30685925594457\\
0.59	0.30215727652338\\
0.6	0.297325840981575\\
0.61	0.292364474430462\\
0.62	0.28727276980134\\
0.63	0.282050388358556\\
0.64	0.276697060197895\\
0.65	0.271212584729776\\
0.66	0.26559683114659\\
0.67	0.25984973887332\\
0.68	0.253971318000351\\
0.69	0.247961649697071\\
0.7	0.241820886604441\\
0.71	0.23554925320421\\
0.72	0.229147046161739\\
0.73	0.22261463463849\\
0.74	0.215952460569027\\
0.75	0.209161038895703\\
0.76	0.20224095775203\\
0.77	0.19519287858263\\
0.78	0.188017536183462\\
0.79	0.180715738640067\\
0.8	0.173288367133127\\
0.81	0.165736375568448\\
0.82	0.158060789970615\\
0.83	0.150262707553026\\
0.84	0.142343295336683\\
0.85	0.134303788127821\\
0.86	0.126145485565962\\
0.87	0.117869747794518\\
0.88	0.109477989040882\\
0.89	0.10097166793869\\
0.9	0.0923522726201031\\
0.91	0.0836212971249371\\
0.92	0.0747802028268292\\
0.93	0.0658303528230594\\
0.94	0.056772894898471\\
0.95	0.04760854029212\\
0.96	0.0383371143056524\\
0.97	0.0289565554694462\\
0.98	0.0194603919237873\\
0.99	0.00983006795750825\\
1	0\\
};
\addlegendentry{Dual bound, $\mu=3$};

\addplot [color=orange,solid]
  table[row sep=crcr]{%
0	0.40568581290711\\
0.01	0.405634501579763\\
0.02	0.403683583034176\\
0.03	0.403788992589408\\
0.04	0.406726635995211\\
0.05	0.404760176387554\\
0.06	0.404934247721604\\
0.07	0.404133432906006\\
0.08	0.404785357246042\\
0.09	0.404657278749659\\
0.1	0.403178195497576\\
0.11	0.402846681141127\\
0.12	0.401681393850156\\
0.13	0.401608671444654\\
0.14	0.400003313824916\\
0.15	0.399780331607568\\
0.16	0.399066333707569\\
0.17	0.400156945336511\\
0.18	0.398526087877023\\
0.19	0.397588322759633\\
0.2	0.397244168090743\\
0.21	0.394488041266189\\
0.22	0.393775326884975\\
0.23	0.392378952888253\\
0.24	0.39211223817025\\
0.25	0.390156062678363\\
0.26	0.388910631830591\\
0.27	0.38733698017507\\
0.28	0.386461766509922\\
0.29	0.383654015531832\\
0.3	0.381591724078295\\
0.31	0.381256725884663\\
0.32	0.378779979106414\\
0.33	0.3767750375301\\
0.34	0.375313146652902\\
0.35	0.373321884471694\\
0.36	0.371469491106711\\
0.37	0.368879514474427\\
0.38	0.36672278024929\\
0.39	0.363978179784419\\
0.4	0.361377526473365\\
0.41	0.358862127861837\\
0.42	0.356016395633893\\
0.43	0.353587619347961\\
0.44	0.350661136449827\\
0.45	0.347779860764171\\
0.46	0.344623573114334\\
0.47	0.341582835416246\\
0.48	0.338004180952541\\
0.49	0.334758858178636\\
0.5	0.331334006107104\\
0.51	0.327849054276787\\
0.52	0.324114229930463\\
0.53	0.320081929986647\\
0.54	0.316172571011846\\
0.55	0.31216550533231\\
0.56	0.307218663164988\\
0.57	0.302885718358606\\
0.58	0.299466225133761\\
0.59	0.293960138788243\\
0.6	0.290736658046101\\
0.61	0.2852677629666\\
0.62	0.281613347898244\\
0.63	0.277395155059028\\
0.64	0.270054492641619\\
0.65	0.26467158225346\\
0.66	0.258924625535442\\
0.67	0.254564118926418\\
0.68	0.247688692322752\\
0.69	0.242275340811037\\
0.7	0.239488886435288\\
0.71	0.231216286676059\\
0.72	0.223746057896451\\
0.73	0.217438634027873\\
0.74	0.214123133526818\\
0.75	0.20515767694773\\
0.76	0.200237341930999\\
0.77	0.193319782382259\\
0.78	0.187087056690234\\
0.79	0.177983637808381\\
0.8	0.172452189166456\\
0.81	0.162448619743193\\
0.82	0.156890074835179\\
0.83	0.146804547800367\\
0.84	0.140845005734264\\
0.85	0.130687832411719\\
0.86	0.124199723604451\\
0.87	0.11745493339922\\
0.88	0.108487847090905\\
0.89	0.102351174866668\\
0.9	0.0907539753646124\\
0.91	0.0820358668977846\\
0.92	0.0741362122455281\\
0.93	0.0673606540627433\\
0.94	0.056981400485814\\
0.95	0.0467837370328542\\
0.96	0.0389703187736889\\
0.97	0.0285538904560352\\
0.98	0.0179629635237942\\
0.99	0.0104436763739154\\
1	6.36460441066801e-18\\
};
\addlegendentry{Achievable};

\end{axis}
\end{tikzpicture}%
  \caption{Capacity upper bounds for $(1,2)$-constrained BEC$(\epsilon)$.}
  \label{fig:bec12}
\end{figure}
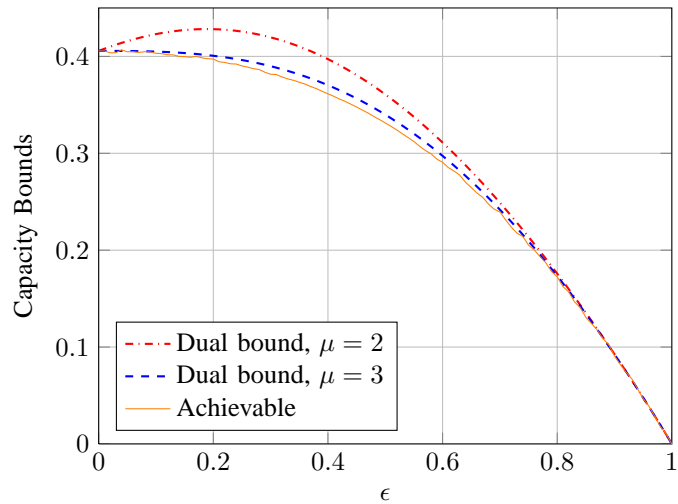

Next, we provide dual bounds for the $(1,2)$-BEC$(\e)$ using $\mu=2$
and $\mu=3$ Markov test distributions.
\begin{theorem}
  (1) The capacity of the $(1,2)$-BEC$(\e)$ satisfies the upper bound
  \begin{align}
&C_{1,2}(\e)\le\dfrac{1-\e}{2}\left[(1-\e)^2\log_2\frac{1}{\b}+\e(1-\e)\log_2\frac{(2-\b)^2}{\b}+\e^2\log_2\frac{(3-\b)^2}{2-\b}\right],\nonumber\\
&\b\in(0,1]\text{ solves }\b^{3(1-\e)}(2-\b)^{2\e-3\e^2}=(1-\b)^{2+2\e-4\e^2}.\nonumber
  \end{align}
(2) The capacity of the $(1,2)$-BEC$(\e)$ satisfies the upper bound
\begin{align}
  &C_{1,2}(\e)\le\dfrac{1-\e}{2}\biggl[(1+\e-2\e^2)\log_2\frac{1}{\b}+2\e^3\log_2(3-\b)+\e^2(3-4\e)\log_2(2-\b)\biggr],\nonumber\\
&\qquad\qquad\qquad\qquad \b\in(0,1]\text{ solves}\nonumber\\
&\b^{3(1+\e-2\e^2)}(2-\b)^{\e^2(3-4\e)}=(1-\b)^{2(1+\e+\e^2-3\e^3)}2^{4\e^2(1-\e)}.\nonumber 
\end{align}
\label{thm:bec12}
\end{theorem}
Fig. \ref{fig:bec12} shows plots of the dual bounds from Theorem
\ref{thm:bec12} along with an achievable rate computed using the simulation
method of \cite{1661831}. 

We conclude the results for the BEC with an upper bound for the
$(d,\infty)$-BEC$(\e)$. Recall the notation $\overline{x}=1-x$ for a
fraction $x$.
\begin{theorem}
  The capacity of the $(d,\infty)$-BEC$(\e)$ satisfies the
 upper bound
\begin{align}
  C_{(d,\infty)}(\e)\le (1-\e)\sum_{i=0}^dB_\e(d,i)\log_2\frac{1+i\overline{\a_0}}{\a_0+i\overline{\a_0}},
\end{align}
where $B_\e(d,i)=\binom{d}{i}\e^i(1-\e)^{d-i}$ and $\a_0\in[0,1]$ solves
\begin{align}
  \sum_{i=0}^dB_\e(d,i)\log_2\frac{(\a_0+i\overline{\a_0})^{d-i+1}}{(1+i\overline{\a_0})^{d-i}}=\log_2\overline{\a_0}.
\end{align}
\label{thm:becd}
\end{theorem}

\subsection{Binary Symmetric Channel}
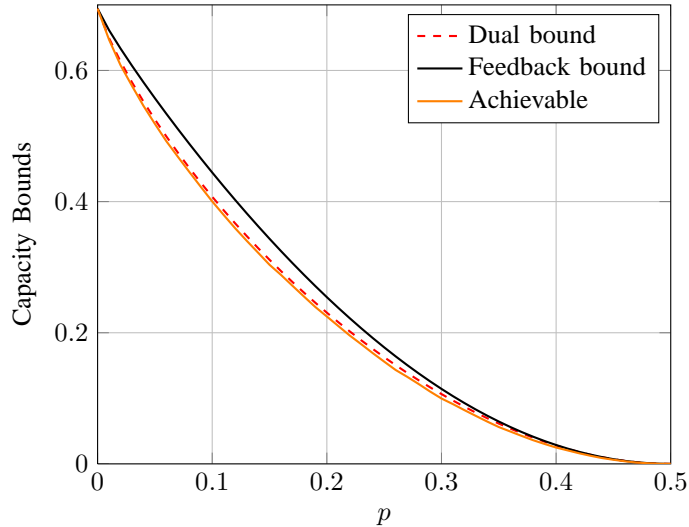
\begin{figure}[htb!]
  \centering
%
%
%
\definecolor{mycolor1}{rgb}{0.00000,0.44700,0.74100}%
\begin{tikzpicture}

\begin{axis}[%
width=3in,
height=2.40441590980269in,
scale only axis,
xmin=0,
xmax=0.5,
xlabel={$p$},
xmajorgrids,
ymin=0,
ymax=0.7,
ylabel={Capacity Bounds},
ymajorgrids,
legend style={draw=black,fill=white,legend cell align=left}
]
\addplot [color=red,dashed,thick]
  table[row sep=crcr]{%
0	0.694241913630617\\
0.01	0.649232508820626\\
0.02	0.613710444815295\\
0.03	0.58192808194695\\
0.04	0.552635216987522\\
0.05	0.525232667779196\\
0.06	0.499365675934317\\
0.07	0.474799190857991\\
0.08	0.451366028429911\\
0.09	0.428941311273383\\
0.1	0.407428370186486\\
0.11	0.386750318139933\\
0.12	0.366844700896477\\
0.13	0.34765993674914\\
0.14	0.329152855592563\\
0.15	0.311286944425615\\
0.16	0.294031064033877\\
0.17	0.277358490025482\\
0.18	0.261246183312064\\
0.19	0.245674226817647\\
0.2	0.230625385199969\\
0.21	0.216084757364291\\
0.22	0.202039500209821\\
0.23	0.188478607948705\\
0.24	0.175392735433824\\
0.25	0.162774056820473\\
0.26	0.150616152947804\\
0.27	0.138913922304077\\
0.28	0.127663511495848\\
0.29	0.116862261880059\\
0.3	0.106508669507244\\
0.31	0.0966023558054701\\
0.32	0.0871440465315451\\
0.33	0.078135556438955\\
0.34	0.0695797768628687\\
0.35	0.0614806629998755\\
0.36	0.0538432170649876\\
0.37	0.0466734627538704\\
0.38	0.0399784055619742\\
0.39	0.0337659725945316\\
0.4	0.0280449246871952\\
0.41	0.0228247331786614\\
0.42	0.0181154138688552\\
0.43	0.0139273119878855\\
0.44	0.0102708348593069\\
0.45	0.00715613374463394\\
0.46	0.0045927431849514\\
0.47	0.00258919453355323\\
0.48	0.00115262906073244\\
0.49	0.000288443015806677\\
0.5	0\\
};
\addlegendentry{Dual bound};

\addplot [color=black,solid,thick]
  table[row sep=crcr]{%
0	0.694241911076359\\
0.01	0.661406436142459\\
0.02	0.633495762319746\\
0.03	0.607256734836707\\
0.04	0.582071504550894\\
0.05	0.557677636610422\\
0.06	0.533939164263871\\
0.07	0.510778705710801\\
0.08	0.48815014259345\\
0.09	0.466025570420765\\
0.1	0.44438833982814\\
0.11	0.423229061003306\\
0.12	0.402543161049314\\
0.13	0.382329300118908\\
0.14	0.362588353176754\\
0.15	0.343322720430531\\
0.16	0.32453580359441\\
0.17	0.306231707308796\\
0.18	0.288414947739685\\
0.19	0.271090317638172\\
0.2	0.254262735055171\\
0.21	0.237937149303938\\
0.22	0.222118481159124\\
0.23	0.206811561964712\\
0.24	0.192021110633682\\
0.25	0.177751700950555\\
0.26	0.164007725416166\\
0.27	0.150793436203408\\
0.28	0.138112858691868\\
0.29	0.125969853954728\\
0.3	0.114368098358454\\
0.31	0.103311053179374\\
0.32	0.09280200348981\\
0.33	0.082844043002768\\
0.34	0.0734400606993135\\
0.35	0.064592787488268\\
0.36	0.0563047505956707\\
0.37	0.0485783120530722\\
0.38	0.0414156354748346\\
0.39	0.0348187358604034\\
0.4	0.0287894356448385\\
0.41	0.0233294079293641\\
0.42	0.0184401374697566\\
0.43	0.0141229712988425\\
0.44	0.0103790783756309\\
0.45	0.00720945880862933\\
0.46	0.00461497791427057\\
0.47	0.00259633145392812\\
0.48	0.00115405020505333\\
0.49	0.000288529389441239\\
0.5	0\\
};
\addlegendentry{Feedback bound};

\addplot [color=orange,solid,thick]
  table[row sep=crcr]{%
0	0.694241333048552\\
0.01	0.646940723745241\\
0.02	0.608076752817161\\
0.03	0.576505727823676\\
0.04	0.546115526367111\\
0.05	0.518566332249006\\
0.06	0.49206003151431\\
0.07	0.468473547355908\\
0.08	0.445289305989757\\
0.09	0.422602710559027\\
0.1	0.40029501537429\\
0.11	0.379622989300219\\
0.12	0.359386622475121\\
0.13	0.340451793588314\\
0.14	0.322111531935922\\
0.15	0.303649771628655\\
0.16	0.288210522104498\\
0.17	0.271672642703301\\
0.18	0.255288416845809\\
0.19	0.239186573647401\\
0.2	0.224775581971485\\
0.21	0.210235557442179\\
0.22	0.195755766268657\\
0.23	0.182528277432902\\
0.24	0.168986956658021\\
0.25	0.155985762480008\\
0.26	0.14322301030857\\
0.27	0.132903278813504\\
0.28	0.121920399345155\\
0.29	0.110407345822871\\
0.3	0.0997077020626397\\
0.31	0.0911409030296691\\
0.32	0.0819827779110219\\
0.33	0.0731760140683375\\
0.34	0.0644828146002328\\
0.35	0.0562042477701774\\
0.36	0.0494209076240213\\
0.37	0.0426291076618645\\
0.38	0.0361024410563796\\
0.39	0.0303619982487252\\
0.4	0.0249490584363963\\
0.41	0.0205990833002048\\
0.42	0.0162356988457076\\
0.43	0.0123758459614\\
0.44	0.00892262971849617\\
0.45	0.00630372548648772\\
0.46	0.00397068366062997\\
0.47	0.00206189923040762\\
0.48	0.000948457495138499\\
0.49	0.000217234748334183\\
0.5	7.91811061162662e-12\\
};
\addlegendentry{Achievable};
\end{axis}
\end{tikzpicture}%
  \caption{Capacity bounds for $(1,\infty)$-constrained BSC$(p)$.}
  \label{fig:bsc}
\end{figure}

The next theorem presents an upper bound to the capacity of the
$(1,\infty)$-BSC$(p)$. We use the notation $H_2(p)=-p\log_2p-(1-p)\log_2(1-p)$.
\begin{theorem}
The capacity of the $(1,\infty)$-BSC$(p)$ satisfies the upper bound
\begin{align}
C_{(1,\infty)}(p)\le\op^2\log_2\frac{1}{a^*}+p\op\log_2\frac{c_1}{\op^2}+p^2\log_2\frac{c_1}{pc_2}-H_2(p),
\end{align}
where $c_1=\overline{a^*}-p^2$, $c_2=2\op-a^*(2-p)$, and $a^*\in[0,1]$ solves
\begin{align}
  a^{2\op}(\ova-p^2)=p^{2p}\op^{2(1-2p)}\ova^{2(1-2p)}[2\op-a(2-p)]^{2p}.
\end{align}
\label{thm:bsc}
  \end{theorem}
The dual bound is plotted along with a feedback-based
bound (extension by authors of
\cite{7308065}) and an achievable rate using the simulation-based
method of \cite{1661831} for comparison in Fig. \ref{fig:bsc}.
We see that the dual capacity bound improves significantly over the
feedback-based bound, and is close to the achievable rate as seen in
the inset.

\subsection{Binary-input AWGN Channel}
Let the Gaussian PDF with mean $\mu$
and variance $\sigma^2$ be denoted
\begin{equation}
  \psi_{\mu,\sigma}(x)\triangleq \frac{1}{\sqrt{2\pi}\sigma}e^{-(x-\mu)^2/2\sigma^2},\,x\in(-\infty,\infty).
\end{equation}
The Gaussian distribution restricted to the interval $(a,b)$ has the PDF 
$\psi_{\mu,\sigma}(x)/\Psi_{\mu,\sigma}(a,b)$, where
$\Psi_{\mu,\sigma}(a,b)\triangleq\int_{a}^b\psi_{\mu,\sigma}(x)dx$ is the
probability that a Gaussian random variable with mean $\mu$ and
variance $\sigma^2$ falls in the interval $(a,b)$.

Consider the binary-input AWGN channel, denoted BIAWGN$(\sigma^2)$,
defined by the relationship $Y=(-1)^X+Z$, where the input $X\in\{0,1\}$
and $Z\sim N(0,\sigma^2)$. The output alphabet $\cY$ is the set of
real numbers. A memory-1 Markov distribution on the output alphabet is specified by
the conditional probability density function (PDF) $q(y_2|y_1)$, where
$y_1$ and $y_2$ are real-valued variables.  

\begin{figure}[htb!]
  \centering
  \begin{tikzpicture}[xscale=1,yscale=1]
\draw[<->] (-4.3,0) -- (4.3,0);
\draw[<->] (0,3) -- (0,-0.5);
\node at (4.3,0) [anchor=north] {$y_2$};
\node at (0,3) [anchor=west] {$q(y_2|y_1)$};
\draw[domain=-1.2:0.5,thick] plot (\x,1);
\draw[domain=-4:-1.2,thick] plot (\x,{2.72*exp(-(\x+2)*(\x+2))});
\draw[domain=0.5:4,thick] plot (\x,{2*exp(-(\x-2)*(\x-2))});
\draw[dashed] (-2,0) -- (-2,2.72);
\draw[dashed] (2,0) -- (2,2);
\draw[dashed] (0.5,0) -- (0.5,1);
\draw[dashed] (-1.2,0) -- (-1.2,1.43);
\draw (2,0.1) -- (2,-0.1);
\draw (-2,0.1) -- (-2,-0.1);
\draw (0.5,0.1) -- (0.5,-0.1);
\draw (-1.2,0.1) -- (-1.2,-0.1);
\node at (2,0) [anchor=north] {\footnotesize  1};
\node at (-2,0) [anchor=north] {\footnotesize -1};
\node at (0.5,0) [anchor=north] {\footnotesize $d_1$};
\node at (-1.2,0) [anchor=north] {\footnotesize $d_2$};
\draw[<->] (-1.2,0.15) -- (0.5,0.15);
\node at (-0.5,0.1) [anchor=south] {\footnotesize $\Delta$};
\node at (-0.5,0.95) [anchor=south] {\footnotesize $b/\Delta$};
\node at (2.2,2) [anchor=west] {\footnotesize $\dfrac{c\,\psi_{1,\sigma}(y_2)}{\Psi_{1,\sigma}(d_1,\infty)}$};
\node at (-2.2,2.72) [anchor=east] {\footnotesize $\dfrac{a\,\psi_{-1,\sigma}(y_2)}{\Psi_{-1,\sigma}(-\infty,d_2)}$};
  \end{tikzpicture}
  \caption{Conditional PDF for a memory-1 Markov distribution.}
\label{fig:qy}
\end{figure}
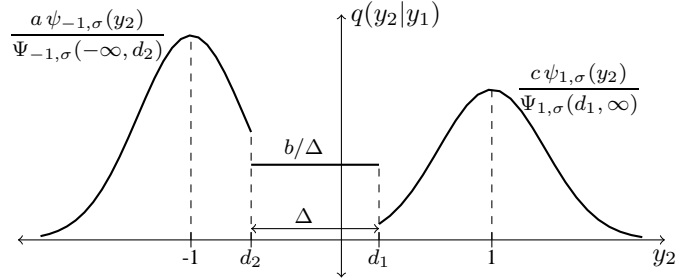

After numerical experimentation, the following form of conditional
PDFs was found to result in tight upper bounds:
\begin{align}
&  q(y_2|y_1)=\nonumber\\
&\begin{cases}
a(y_1,\sigma)\dfrac{\psi_{-1,\sigma}(y_2)}{\Psi_{-1,\sigma}(-\infty,d_2(y_1,\sigma))},\,y_2<d_2(y_1,\sigma),\\[10pt]
b(y_1,\sigma)\dfrac{1}{\Delta(y_1,\sigma)},\quad y_2\in[d_2(y_1,\sigma),d_1(y_1,\sigma)],\\[8pt]
c(y_1,\sigma) \dfrac{\psi_{+1,\sigma}(y_2)}{\Psi_{+1,\sigma}(d_1(y_1,\sigma),+\infty)},\,y_2>d_1(y_1,\sigma),
\end{cases}\label{eq:12}
\end{align}
where $d_1$, $d_2$, $a$, $b$ and $c$ are functions of $y_1$
and $\sigma$, and need to be chosen suitably for validity of the PDF
and for minimizing the upper
bound.  We define $\Delta(y_1,\sigma)\triangleq d_1(y_1,\sigma)-d_2(y_2,\sigma)$. The conditional PDF is illustrated in Fig. \ref{fig:qy}.
In the figure, the dependence on $y_1$ and $\sigma$ in the functions is suppressed to
reduce clutter. The conditional PDF is restricted Gaussian with means
+1 and -1 in the intervals $(d_1,\infty)$ and $(-\infty,d_2)$,
respectively, and uniform in the interval $[d_2,d_1]$. The
values of $a$, $b$, and $c$ are fractions summing to 1 for each value
of $y_1$ and $\sigma$, and this constraint makes $q(y_2|y_1)$ a
valid PDF for every $y_1$.

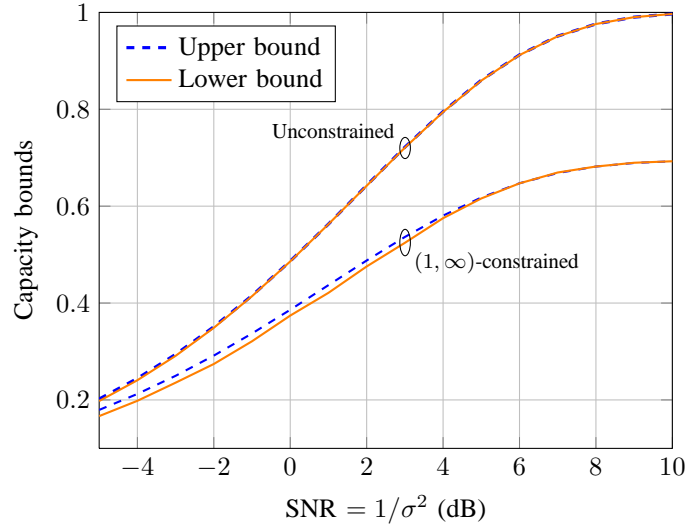
\begin{figure}[htb!]
  \centering
%
%
%
\definecolor{mycolor1}{rgb}{0.00000,0.44700,0.74100}%
\definecolor{mycolor2}{rgb}{0.85000,0.32500,0.09800}%
\definecolor{mycolor3}{rgb}{0.92900,0.69400,0.12500}%
\definecolor{mycolor4}{rgb}{0.49400,0.18400,0.55600}%
\begin{tikzpicture}

\begin{axis}[%
width=3in,
height=2.28372652864709in,
scale only axis,
xmin=-5,
xmax=10,
xlabel={SNR $=1/\sigma^2$ (dB)},
xmajorgrids,
ymin=0.1,
ymax=1,
ylabel={Capacity bounds},
ymajorgrids,
legend style={draw=black,fill=white,legend cell align=left,legend pos=north west}
]
\addplot [color=blue,dashed,very thick]
  table[row sep=crcr]{%
-5	0.2019\\
-4	0.2444\\
-3	0.294\\
-2	0.351\\
-1	0.4154\\
0	0.4865\\
1	0.563\\
2	0.6429\\
3	0.7217\\
4	0.7953\\
5	0.86\\
6	0.9124\\
7	0.951\\
8	0.9761\\
9	0.9902\\
10	0.9968\\
};
\addlegendentry{Upper bound};

\addplot [color=orange,solid,thick]
  table[row sep=crcr]{%
-5	0.1977\\
-4	0.2407\\
-3	0.291\\
-2	0.3489\\
-1	0.4141\\
0	0.4859\\
1	0.5628\\
2	0.6421\\
3	0.7207\\
4	0.7944\\
5	0.8592\\
6	0.9119\\
7	0.9507\\
8	0.976\\
9	0.9902\\
10	0.9968\\
};
\addlegendentry{Lower bound};

\addplot [color=blue,dashed,thick]
  table[row sep=crcr]{%
-5	0.1794\\
-4	0.2124\\
-3	0.2498\\
-2	0.2915\\
-1	0.3372\\
0	0.3861\\
1	0.437\\
2	0.4878\\
3	0.5364\\
4	0.5805\\
5	0.6179\\
6	0.6473\\
7	0.6684\\
8	0.6818\\
9	0.6892\\
10	0.6926\\
};

\addplot [color=orange,solid,thick]
  table[row sep=crcr]{%
-5	0.1666\\
-4	0.1982\\
-3	0.2356\\
-2	0.2742\\
-1	0.3211\\
0	0.3741\\
1	0.4211\\
2	0.4755\\
3	0.5243\\
4	0.5752\\
5	0.6159\\
6	0.6471\\
7	0.6695\\
8	0.6818\\
9	0.6896\\
10	0.6927\\
};

\node at (axis cs:3,0.72) (un) {\tikz{\draw ellipse (2pt and 4pt);}};
\node [anchor=south east] at (un) {\footnotesize Unconstrained};

\node at (axis cs:3,0.525) (con) {\tikz{\draw ellipse (2pt and 5pt);}};
\node [anchor=north west] at (con) {\footnotesize $(1,\infty)$-constrained};

\end{axis}
\end{tikzpicture}%
  \caption{Capacity bounds for the unconstrained and
    $(1,\infty)$-constrained BIAWGN$(\sigma^2)$.}
  \label{fig:bpsk}
\end{figure}

By numerical optimization of the functions $d_1$, $\Delta$, $a$ and
$b$ in the test distribution (note that $d_2=d_1-\Delta$ and
$c=1-a-b$), an upper bound can
be obtained for the $(1,\infty)$-constrained
BIAWGN$(\sigma^2)$. The results are shown in Fig. \ref{fig:bpsk}.
In Fig. \ref{fig:bpsk}, a dual capacity upper bound for the
unconstrained BIAWGN channel derived using a test distribution similar to the
one shown in Fig. \ref{fig:bpsk} is
also shown. Details of the computations are provided in later sections. The upper bounds
are seen to be quite tight when compared to lower bounds in both the
constrained and the unconstrained case. The lower bound was computed using the simulation
method of \cite{1661831} for the constrained case. We remark that the choice of the test
distribution is based on simple heuristics and could possibly be improved to
obtain better bounds.

The above examples illustrate the kind of upper bounds that can be derived
using the dual method. These are illustrative and not meant to be
exhaustive. In general, for discrete channels, an algebraic
characterization is possible, while for Gaussian channels, we obtain a
small-scale numerical computation for constrained capacity. The
computational complexity increases with the memory of the Markov
distribution $\mu$, and the optimization problems are nonlinear and may not be convex,
in general. However, since their scale is small (number of variables and
constraints depend on the memory $\mu$), standard computation packages
are effective in solving the optimizations or providing good local minima. The choice of test
distribution can be guided by Lagrangian methods, but may also be
made using heuristics for minimizing the relative entropy in the dual
bound.

\section{Proofs and Computations}
\label{sec:proofs-computations}
In this section, we provide proofs for the theorems in Section
\ref{sec:upper-bounds} and show the computations involved in
simplifying the expressions for upper bounds. Some of the routine
computations are given in the appendix.

We begin with the basic underlying idea, which is the use of the dual capacity upper bound.
\subsection{Dual capacity bound}
Consider the $(d,k)$-constrained channel $p_{Y|X}$. An upper bound on
the $N$-letter capacity, called the dual capacity upper bound \cite{topsoe67}\cite{KEMPERMAN1974101}\cite{Csiszar11}, is given by the following:
\begin{equation}
  \label{eq:2}
  C^N_{d,k}(p_{Y|X})\le \max_{x^N\in\cX_{d,k}^N}D\bigl(p_{Y^N|X^N}(\,\cdot\,|x^N)\|q_{Y^N}(\,\cdot\,)\bigr),
\end{equation}
where $q_{Y^N}$, called the test distribution, is an arbitrary distribution on the output alphabet
$\cY^N$. Note that in the above bound
$x^N$ is fixed in the expression $D\bigl(p_{Y^N|X^N}(\,\cdot\,|x^N)\|q_{Y^N}(\,\cdot\,)\bigr)$, and the
relative entropy is evaluated between the
probability mass functions (PMFs) $p_{Y^N|X^N}(y^N|x^N)$ and $q_{Y^N}(y^N)$ over
the alphabet $\cY^N$. The maximum in \eqref{eq:2} is over all valid $(d,k)$
sequences $x^N$.

Since the bound in \eqref{eq:2} holds for every $q_{Y^N}$, we
minimize over a family of distributions $\cQ$ to improve the upper
bound as follows:
\begin{equation}
  C^N_{d,k}(p_{Y|X})\le \min_{q_{Y^N}\in\cQ}\max_{x^N\in\cX_{d,k}^N}D\bigl(p_{Y^N|X^N}(\,\cdot\,|x^N)\|q_{Y^N}(\,\cdot\,)\bigr).
\label{eq:3}
\end{equation}
Given the nature of the input constraint, a good choice for the family
of test distributions is $\cQ_{\mu}$, which is the family of Markov distributions on $\cY$ with memory $\mu$.
Therefore, an upper bound to the capacity of the $(d,k)$-constrained
$p_{Y|X}$ is 
\begin{align}
&C_{d,k}(p_{Y|X})\le\min_{q_{Y^N}\in\cQ_{\mu}}\limsup_{N\to\infty}\frac{1}{N}\max_{x^N\in\cX_{d,k}^N}D\bigl(p_{Y^N|X^N}(\,\cdot\,|x^N)\|q_{Y^N}(\,\cdot\,)\bigr).
\label{eq:7}
\end{align}
The above bound is not directly computable because of the dependence
on $N$. We simplify the bound by application of the chain rule for
relative entropy, and restrict the test distributions to obtain the
dual bound theorem and corollary. The simplification is described
below in detail for completeness.

\subsection{Simplifying the dual capacity bound}
For $q\in\cQ_{\mu}$ (the subscript $Y^N$ is suppressed in the notation), $x^N\in\cX_{d,k}^N$ and a channel $p_{Y|X}$, the relative entropy term in the dual capacity bound can be simplified as follows:
\begin{align}
&  D\bigl(p_{Y^N|X^N}(\,\cdot\,|x^N)\|q(\,\cdot\,)\bigr)\nonumber\\
&=\sum_{y^N\in\cY^N}p_{Y^N|X^N}(y^N|x^N)\log_2\frac{\prod_{n=1}^Np_{Y|X}(y_n|x_n)}{\prod_{n=1}^{\mu}q(y_n|y^{n-1})\prod_{n=\mu+1}^Nq(y_n|y_{[n-\mu:n-1]})}\nonumber\\
&=\sum_{y^N\in\cY^N}\left(\prod_{i=1}^Np_{Y|X}(y_i|x_i)\right)\biggl[\sum_{n=1}^{\mu}\log_2\frac{p_{Y|X}(y_n|x_n)}{q(y_n|y^{n-1})}+\sum_{n=\mu+1}^N\log_2\frac{p_{Y|X}(y_n|x_n)}{q(y_n|y_{[n-\mu:n-1]})}\biggr]\nonumber\\
&\overset{(a)}{=}\sum_{n=1}^{\mu}\biggl[\,\sum_{y^{n-1}\in\cY^{n-1}}\left(\prod_{i=1}^{n-1}p_{Y|X}(y_i|x_i)\right) \sum_{y_n\in\cY} p_{Y|X}(y_n|x_n)\log_2\frac{p_{Y|X}(y_n|x_n)}{q(y_n|y^{n-1})}\biggr]+\nonumber\\
&\qquad\sum_{n=\mu+1}^N\biggl[\,\,\sum_{y_{[n-\mu:n-1]}\in\cY^{\mu}}\left(\prod_{i=n-\mu}^{n-1}p_{Y|X}(y_i|x_i)\right) \sum_{y_n\in\cY} p_{Y|X}(y_n|x_n)\log_2\frac{p_{Y|X}(y_n|x_n)}{q(y_n|y_{[n-\mu:n-1]})}\biggr],\nonumber\\
&\overset{(b)}{=}\sum_{n=1}^{\mu}\biggl[\,\sum_{y^n\in\cY^n}p_{Y^{n-1}|X^{n-1}}(y^{n-1}|x^{n-1}) D\bigl(p_{Y|X}(\,\cdot\,|x_n)\|q(\,\cdot\,|y^{n-1})\bigr)\biggr]+\nonumber\\
&\qquad\sum_{n=\mu+1}^N\biggl[\,\sum_{y^{\mu}\in\cY^{\mu}}p_{Y^{\mu}|X^{\mu}}(y^{\mu}|x_{[n-\mu:n-1]}) D\bigl(p_{Y|X}(\,\cdot\,|x_n)\|q(\,\cdot\,|y^{\mu})\bigr)\biggr],
  \label{eq:6}
\end{align}
where $(a)$ follows by interchanging the order of summation (over $y^N$
and over $n$) and marginalizing, and $(b)$ follows by identifying the
expressions as relative entropies and changing the dummy summation
variables from $y_{[n-\mu:n-1]}$ to $y^{\mu}$.

The first summation term from $n=1$ to $n=\mu$ in \eqref{eq:6} is
$o(N)$ (assuming that the choice of $q$ is such that the relative
entropy is finite). Since the dual bound (see \eqref{eq:7}) involves division by $N$,
which tends to infinity, any $o(N)$ term is insignificant in the final
bound on $C_{d,k}$. Moreover, the second summation term from $n=\mu+1$
to $n=N$ in \eqref{eq:6} is readily identified as the sum of edge
metrics (see \eqref{eq:10}) of the walk in the graph $G^{\mu}_{d,k}$ corresponding to the $(d,k)$ sequence $x^N$. So, we have
\begin{align}
\frac{1}{N}D\bigl(p_{Y^N|X^N}(\,\cdot\,|x^N)\|q(\,\cdot\,)\bigr)=\frac{1}{N}\sum_{n=\mu+1}^NT_{q,p_{Y|X}}(x_{[n-\mu:n]})+o(1),   \label{eq:11}
\end{align}
where $o(1)$ tends to 0 as $N\to\infty$. Therefore, the edge metrics 
on the walk corresponding to $x^N$ add up to form the asymptotically
significant part of the upper bound on capacity. 

\subsection{Decomposing walks into cycles}
A standard result in graph theory is that any walk in a directed graph
can be decomposed
into a path and a set of cycles (see \cite{bang2013digraphs},
Exercises of Chapter 1). For completeness and future use, we provide a brief algorithmic proof of this
result for the graph $G^{\mu}_{d,k}$.

Like mentioned before, a length-$N$, $(d,k)$-sequence
$x^N\in\cX^N_{d,k}$ corresponds to a walk $w(x^N)$ of length $N-\mu$ with
sequence of vertices $v_i=x_{[i:i+\mu-1]}$ for
$i=1,2,\ldots,N-\mu+1$. In the walk $w(x^N)$, let $i_1$ be the least
integer for which $v_{i_1}$ occurs more than once in $w(x^N)$, and let
$i_2$ be the least integer such that $i_2>i_1$ and
$v_{i_1}=v_{i_2}$. Basically, $c_1(x^N)=(v_{i_1},v_{i_1+1},\ldots,v_{i_2})$ is
the sequence of vertices of the first cycle traversed in $w(x^N)$, and its
length is $l(c_1(x^N))=i_2-i_1$.

Now, remove the edges of $c_1(x^N)$ from the walk $w(x^N)$, and
consider the walk $w_1=(v_1,\ldots,v_{i_1},v_{i_2+1},\ldots)$. The
first cycle in $w_1$ is denoted $c_2(x^N)$, and the same process is
continued iteratively till a path $w^*(x^N)$ without any repeating
vertices results after, say, $n_c(x^N)$ steps. The cycles resulting in
this process $\{c_i(x^N)\}$ for $i=1,2,\ldots,n_c(x^N)$ and the final
path $w^*(x^N)$ form the decomposition of the walk $w(x^N)$ into a set of
cycles and a path in $G^{\mu}_{d,k}$. 

We now use the above decomposition of the walk $w(x^N)$ in the dual
capacity expression in \eqref{eq:11}. Since the metric of the path
$w^*(x^N)$ is $o(N)$, we have that
\begin{align} \frac{1}{N}D\bigl(p_{Y^N|X^N}(\,\cdot\,|x^N)\|q(\,\cdot\,)\bigr)=\frac{1}{N}\sum_{i=1}^{n_c(x^N)}T_{q,p_{Y|X}}(c_i(x^N))+o(1).\label{eq:5}
\end{align}
Since the length of the path
$w^*(x^N)$ is $o(N)$, we have that
\begin{equation}
  \frac{1}{N}\sum_{i=1}^{n_c(x^N)}l(c_i(x^N))=1+o(1).
\label{eq:21}
\end{equation}
The above two equations resulting from the decomposition of walks into
cycles along with a cycle metric restriction on the Markov
distribution result in the proof of dual bound theorem. The
restriction is described next.

\subsection{KKT-constrained Markov test distributions}
It is easy to show that the dual capacity upper bound results in
equality if the test distribution is set to be equal to the
capacity-achieving output distribution. Now, one of the
classic results in computation of capacity of a channel $X\to Y$ with
channel transition probability $p(y|x)$ is the set of
Karush-Kuhn-Tucker (KKT)
conditions, which require that 
\begin{align*}
  D\bigl(p(y|x)\|p(y)\bigr)&=C,\text{ if }p(x)>0,\\
  D\bigl(p(y|x)\|p(y)\bigr)&\le C,\text{ if }p(x)=0,
\end{align*}
for the capacity-achieving output distribution $p(y)$. Therefore, in
the dual capacity bound calculation for the $(d,k)$-constrained
channel $p_{Y|X}$, a simplifying condition that could result in good bounds is to require that the test
distribution $q(y^N)$ satisfies
\begin{align}
  \frac{1}{N}D\bigl(p_{Y^N|X^N}(\,\cdot\,|x^N)\|q(\,\cdot\,)\bigr)=\text{constant}+o(1),\text{ if }x^N\in\cX^N_{d,k},\label{eq:13}
\end{align}
where the constant is independent of $x^N$, but possibly
dependent on $q$ and $p_{Y|X}$. 

Consider the restricted set of Markov test distributions
\begin{equation}
  \cQ^*_{\mu}\triangleq\{q\in\cQ_{\mu}:\frac{T_{q,p_{Y|X}}(c)}{l(c)}=\frac{T_{q,p_{Y|X}}(c')}{l(c')}\forall
  c,c'\in C^{\mu}_{d,k}\}.
\label{eq:23}
\end{equation}
In words, for $q\in\cQ^*_{\mu}$, the length-normalized metric of every
cycle in $G^{\mu}_{d,k}$ is a constant. The common value of the
length-normalized cycle metric is denoted $t(q,p_{Y|X})$, i.e., 
\begin{equation}
  t(q,p_{Y|X})\triangleq\frac{T_{q,p_{Y|X}}(c)}{l(c)}, q\in\cQ^*_{\mu}, c\in C^{\mu}_{d,k}.
\end{equation}
Using the cycle decomposition of walks
and the results in \eqref{eq:5}, \eqref{eq:21}, we readily see that,
for $q\in\cQ^*_{\mu}$ and $x^N\in\cX^N_{d,k}$,
\begin{align}
  \frac{1}{N}D\bigl(p_{Y^N|X^N}(\,\cdot\,|x^N)\|q(\,\cdot\,)\bigr)&=\frac{1}{N}\sum_{i=1}^{n_c(x^N)}T_{q,p_{Y|X}}(c_i(x^N))+o(1)\nonumber\\
&=t(q,p_{Y|X})\left(\frac{1}{N}\sum_{i=1}^{n_c(x^N)}l(c_i(x^N))\right)+o(1)\nonumber\\
&=t(q,p_{Y|X})+o(1).\label{eq:24}
\end{align}
We will refer to the set of distributions $\cQ^*_{\mu}$ as the
KKT-constrained Markov test distributions of memory $\mu$. 
This is to be contrasted with the choice of test distributions made
using the maximizing branch transition probability method in \cite{955166}.
\subsection{Proof of Theorem \ref{thm:ubq} and Corollary \ref{thm:ubQ}}
The proof of the dual bound theorem and corollary is now
immediate. Using \eqref{eq:24}, we see that
\begin{align}
\limsup_{N\to\infty}\frac{1}{N}\max_{x^N\in\cX_{d,k}^N}&D\bigl(p_{Y^N|X^N}(\,\cdot\,|x^N)\|q_{Y^N}(\,\cdot\,)\bigr)\nonumber\\
&=t(q,p_{Y|X}).
\end{align}
Hence, the dual capacity bound for a KKT-constrained Markov test
distribution results in
\begin{equation}
  C_{d,k}(p_{Y|X})\le t(q,p_{Y|X}),\quad q\in\cQ^*_{\mu}, 
\end{equation}
which proves Theorem \ref{thm:ubq}. Minimization over
$q\in\cQ^*_{\mu}$ proves Corollary \ref{thm:ubQ}.

\subsection{Binary Erasure Channel}
\subsubsection{$(1,\infty)$-BEC$(\e)$ (Proof of Theorem \ref{thm:bec} - Part (1))}
For this bound, the dual bound corollary is used with a memory-1
Markov test distribution given in \eqref{eq:4a}, which is reproduced here for convenience.
\begin{equation}
   q(y_2|y_1)=\begin{bmatrix}
\b(1-\e)&\e&\ob(1-\e)\\
\a(1-\e)&\e&\oa(1-\e)\\
1-\e&\e&0
\end{bmatrix},
   \label{eq:4}
\end{equation}
where the rows correspond to $y_1=0,?,1$, the columns correspond to $y_2=0,?,1$
in that order ($?$ denotes erasure symbol), and $\a,\b$ are parameters satisfying
$0\le\a,\b\le1$. The specific choice in \eqref{eq:4} is motivated by the minimization
involved in the upper bound. If $y_{n-1}=1$, we
have $x_{n-1}=1$ because the channel is a BEC. Now, if
$x_{n-1}=1$, by the $(1,\infty)$ constraint, we have that $x_n=0$. So,
$q(y_2|1)$ can be set to be $p_{Y|X}(y_2|0)=[1-\e\,\,\e\,\,0]$ to ensure that
$D\bigl(p_{Y|X}(\,\cdot\,|x_n)\|q(\,\cdot\,|y_{n-1})\bigr)=0$ whenever $y_{n-1}=1$. The choice of
$q(?|y_1)=\e$ can be shown to be best possible for the BEC by using
complementary slackness in the optimization problem of Corollary
\ref{thm:ubQ}, and we skip the details. 

With the choice of $q(y_2|y_1)$ as in \eqref{eq:4}, the metrics
$T(00)$, $T(01)$ and $T(10)$ were computed earlier and are given in
\eqref{eq:26}, \eqref{eq:27} and \eqref{eq:28}. We will drop the
subscripts from the notation $T_{q,p_{Y|X}}(\,\cdot\,)$ to reduce
clutter. Note that the metrics depend on the parameters $\a$, $\b$ and
the channel erasure probability $\e$.

As seen in Fig. \ref{fig:sdmem}(a), there are two cycles in $G^1_{1,\infty}$ - a length-1 cycle associated
to the sequence $(00)$, and a length-2 cycle associated to the
sequence $(010)$. The length-normalized cycle metrics are $T(00)$ and
$(T(01)+T(10))/2$, which have to be equal for a KKT-constrained
Markov test distribution. So, we have the constraint $T(00)=(T(01)+T(10)/2$.
By the dual bound corollary, we get the bound
\begin{align}
C_{1,\infty}(\e)\le\min_{\substack{q:\\T(01)+T(10)=2T(00)}}T(00).
\label{eq:22}
\end{align}
The above constrained optimization can be solved by standard
Lagrangian techniques (as shown in the Appendix) resulting in the
upper bound in Theorem \ref{thm:bec} - Part (1). 

In the optimization problem of \eqref{eq:22} and those in the ensuing
examples, while our methods might be resulting in global minima, we
have not made an attempt to prove the same. The optimization problems
could be non-convex in some cases, but Lagrangian methods result in
good bounds even though there is a chance that they might be local minima for the problems. 

\subsubsection{$(1,\infty)$-BEC$(\e)$ (Proof of Theorem \ref{thm:bec} - Part (2))}
For this bound, the dual bound corollary is used with a memory-2
Markov test distribution shown in Table \ref{tab:m2}. 
\begin{table}[htb]
  \centering
  \caption{Choice of $q(y_3|y_1,y_2)$, $(rs)\in\{00,0?,?0,??,10,1?\}$.}
  \begin{tabular}{|c|c|}
\hline
    $(y_1y_2)$&$q(y_3|y_1,y_2)$\\\hline
01, ?1&$[1-\e\,\,\,\,\e\quad\,0\quad]$\\\hline
$rs$&$[\a_{rs}(1-\e)\,\,\e\,\,\oa_{rs}(1-\e)]$\\
\hline
  \end{tabular}
\label{tab:m2}
\end{table}
The choices for $q(y_3|y_1,y_2)$ have been made to suit the constraint
and the channel. If $y_2=1$, then we have $x_2=1$ since the channel
is a BEC. This implies that $x_3=0$. So, we set
$q(y_3|y_1,1)=p_{Y|X}(y|0)$. For other values of $y_1=r$, $y_2=s$,
$q(y_3|y_1,y_2)$ is parameterized by $\a_{rs}\in[0,1]$. 

As seen from Fig. \ref{fig:sdmem}(b), there are three directed
cycles in the state diagram $G^2_{1,\infty}$ - (1) length-1 cycle associated to
$(000)$, (2) length-2 cycle associated to $(0101)$, and (3) length-3
cycle associated to $(00100)$. Equating the length-normalized cycle metrics, we get the constraints
\begin{align*}
  T(010)+T(101)&=2T(000),\\
T(001)+T(010)+T(100)&=3T(000).
\end{align*}
Note that the metrics $T(\,\cdot\,)$ are functions of the parameters
$\a_{rs}$ and the erasure probability $\e$. By the dual bound corollary, we get the bound
\begin{align}
C_{(1,\infty)}(\e)\le
\min_{\substack{q:\\T(010)+T(101)=2T(000)\\T(001)+T(010)+T(100)=3T(000)}}T(000).
\label{eq:25}
\end{align}
The above constrained optimization can be solved by standard
Lagrangian techniques (as shown in the Appendix) resulting in the
upper bound in Theorem \ref{thm:bec} - Part (2). 

\subsubsection{$(1,2)$-BEC$(\e)$ (Proof of Theorem \ref{thm:bec12})}
\begin{figure*}[htb]
  \centering
  \tikzstyle{cnode}=[circle,draw]
  \begin{tikzpicture}
    \begin{scope}[node distance=3cm,>=angle 90,semithick]
    \node[cnode] (n00)                   {00};
    \node[cnode] (n01) [right of=n1,xshift=3cm]     {01};
    \node[cnode] (n10) [below of=n1,xshift=3cm]     {10};   
    \draw[->] (n00) -- node[above]{$T(001)$} node[below]{1} (n01);
    \draw[->] (n10) -- node[below left]{$T(100)$} node[above right]{0} (n00);
    \draw[->] (n10) to[out=70,in=200] node[left]{$T(101)$} node[right]{1} (n01);
    \draw[->] (n01) to[out=250,in=20] node[left]{0} node[right]{$T(010)$} (n10);
    \end{scope}
    \node[anchor=west] at (0,-3.7) {\footnotesize (a) $(1,2)$-constrained, $\mu=2$.};
        \begin{scope}[shift={(8.5,0)},node distance=3cm,>=angle 90,semithick]
    \node[cnode] (n001)                   {001};
    \node[cnode] (n010) [right of=n001,xshift=2cm]     {010};
    \node[cnode] (n100) [below of=n001,xshift=2.5cm]     {100};
    \node[cnode] (n101) [right of=n100,xshift=2cm]     {101};
    \draw[->] (n001) -- node[above]{$T(0010)$} node[below]{0} (n010);
    \draw[->] (n010) -- node[left]{$T(0100)$} node[right]{0} (n100);
    \draw[->] (n100) -- node[left]{$T(1001)$} node[right]{1} (n001);
    \draw[->] (n010) to[out=350,in=100] node[right]{$T(0101)$}
    node[below left]{1} (n101);
    \draw[->] (n101) to[out=170,in=280] node[left]{$T(1010)$}
    node[above right]{0} (n010);
    \end{scope}
    \node[anchor=west] at (9,-3.7) {\footnotesize (b) $(1,2)$-constrained, $\mu=3$.};
  \end{tikzpicture} 
\caption{$G^{\mu}_{1,2}$: Memory-$\mu$ state diagram for the $(1,2)$ constraint.}
\label{fig:sdmem12}
\end{figure*}
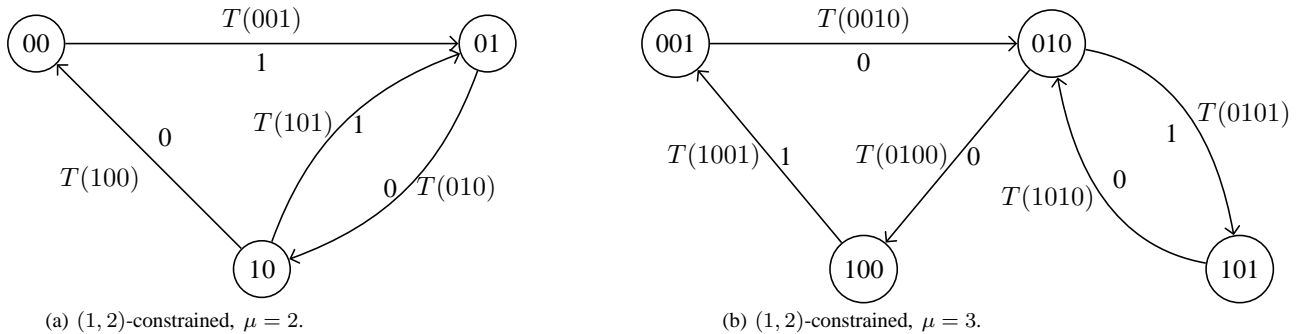

The state diagrams for the $(1,2)$ constraint with $\mu=2,3$ are shown
in Fig. \ref{fig:sdmem12}.
The Markov test distribution used for $\mu=2$ is given in Table \ref{tab:test12mu2}.
\begin{table}[htb]
  \centering
  \caption{Test distribution family for $\mu=2$. $(rs)\in\{10,0?,?0,1?,??\}$.}
  \begin{tabular}{|c|c|}
\hline
    $(y_1y_2)$&$q(y_3|y_1,y_2)$\\\hline
00&$[0\qquad\,\,\,\e\,\,1-\e]$\\\hline
01, ?1&$[1-\e\,\,\,\,\e\quad\,0\quad]$\\\hline
$rs$&$[\a_{rs}(1-\e)\,\,\e\,\,\oa_{rs}(1-\e)]$\\
\hline
  \end{tabular}
\label{tab:test12mu2}
\end{table}
The choices are made to suit the constraint. For $(y_1y_2)\in\{01,?1\}$, we have $x_3=0$. So, the test distribution $q(y_3|0,1)$ and
$q(y_3|?,1)$ are both made equal to
$p_{Y|X}(\,\cdot\,|0)=[1-\e\,\,\e\,\,0]$. Likewise, when
$(y_1y_2)=00$, we have $x_3=1$ by the $(1,2)$ constraint. So, we set
$q(y_3|0,0)$ as $p_{Y|X}(\,\cdot\,|1)=[0\,\,\e\,\,1-\e]$. For other
possibilities, we introduce parameters $\a_{rs}\in[0,1]$.

In $G^{2}_{1,2}$, as seen in Fig. \ref{fig:sdmem12}(a), there are two
cycles - (1) the length-2 cycle associated to $(0101)$, and (2) the
length-3 cycle associated to $(00100)$. Equating the length-normalized
cycle metrics, we get the constraint
\begin{align}
  \frac{T(010)+T(101)}{2}=\frac{T(001)+T(010)+T(100)}{3}.
\end{align}
Note that the metrics $T(\,\cdot\,)$ are functions of the parameters
$\a_{rs}$ and the erasure probability $\e$. By the dual bound
corollary, we get the bound
\begin{align}
C_{(1,2)}(\e)\le
\min_{\substack{q:\\T(010)+T(101)=2t\\T(001)+T(010)+T(100)=3t}}t.
\label{eq:29}
\end{align}
The above constrained optimization can be solved by standard
Lagrangian techniques (as shown in the Appendix) resulting in the
upper bound in Theorem \ref{thm:bec12} - Part (1). 

The Markov test distribution used for $\mu=3$ is given in Table \ref{tab:test12mu3}.
\begin{table}[htb]
  \centering
  \caption{Test distribution family for $\mu=3$. $(rst)\in\{10?,0??,?0?,??0,1??,???\}$.}
  \begin{tabular}{|c|c|}
\hline
    $(y_1y_2y_3)$&$q(y_4|y_1,y_2,y_3)$\\\hline
1?0, ?00, 100&$[0\qquad\,\,\,\e\,\,1-\e]$\\\hline
??1, 0?1, 00?, 001, 1?1, ?01, 101&$[1-\e\,\,\,\,\e\quad\,0\quad]$\\\hline
?10, 01?, 0?0, ?1?, 010&$[\a_{010}(1-\e)\,\,\e\,\,\oa_{010}(1-\e)]$\\\hline
$rst$&$[\a_{rst}(1-\e)\,\,\e\,\,\oa_{rst}(1-\e)]$\\
\hline
  \end{tabular}
\label{tab:test12mu3}
\end{table}
The choices are made to suit the constraint. For
$(y_1y_2y_3)\in\{1?0,?00,100\}$, we have $x_4=1$. So, the conditional test
distribution for these cases is set as 
$p_{Y|X}(\,\cdot\,|1)=[0\,\,\e\,\,1-\e]$. Likewise, when
$y_3=1$ or $(y_1y_2y_3)=00?$, we have $x_4=0$ by the $(1,2)$ constraint. So, we set
the conditional test distribution for these cases as $p_{Y|X}(\,\cdot\,|0)=[1-\e\,\,\e\,\,0]$.
For $(y_1y_2y_3)\in\{?10,01?,0?0,?1?,010\}$, we see that
$(x_1x_2x_3)=010$. So, we set the conditional test distribution for
these cases as $[\a_{010}(1-\e)\,\,\e\,\,\oa_{010}(1-\e)]$ with a
common parameter $\a_{010}$. For every other possibility, a parameter
$\a_{rst}\in[0,1]$ is used.

In $G^{3}_{1,2}$, as seen in Fig. \ref{fig:sdmem12}(b), there are two
cycles - (1) the length-2 cycle associated to $(01010)$, and (2) the
length-3 cycle associated to $(001001)$. Equating the length-normalized
cycle metrics, we get the constraint
\begin{align}
  \frac{T(0101)+T(1010)}{2}=\frac{T(0010)+T(0100)+T(1001)}{3}.
\end{align}
Note that the metrics $T(\,\cdot\,)$ are functions of the parameters
$\a_{rst}$ and the erasure probability $\e$. By the dual bound
corollary, we get the bound
\begin{align}
C_{(1,2)}(\e)\le
\min_{\substack{q:\\T(0101)+T(1010)=2t\\T(0010)+T(0100)+T(1001)=3t}}t.
\label{eq:30}
\end{align}
The above constrained optimization can be solved by standard
Lagrangian techniques (as shown in the Appendix) resulting in the
upper bound in Theorem \ref{thm:bec12} - Part (2). 

\subsubsection{$(d,\infty)$-BEC$(\e)$ (Proof of Theorem \ref{thm:becd})}
For the $(d,\infty)$ constraint with a test distribution of memory
$\mu=d$, the graph $G^{d}_{d,\infty}$ has $d+1$ states and is
isomorphic to the graph shown in
Fig. \ref{fig:dk}(b) with the labels of states replaced with
$(d,\infty)$-constrained sequences of length $d$. The state labels in
$G^{d}_{d,\infty}$ are the $d+1$ sequences in
$\cX^{d}_{d,\infty}=\{u_0,u_1,\ldots,u_d\}$, where $u_0$ is the
all-zero sequence of length $d$ and $u_i$ is the length-$d$
binary sequence with a single 1 in the $i$-th position for $1\le i\le d$.

For the bound in Theorem \ref{thm:becd}, the dual bound corollary is
used with the memory-$d$ Markov test
distribution $q(y_{d+1}|y^d)$ defined as follows:
\begin{align}
  q(y_{d+1}|y^d)=\begin{cases}
[1-\e\ \e\ 0]&\text{ if }y^d\text{ has a }1,\\
[\a_0(1-\e)\ \e\ \overline{\a_0}\,\overline{\e}]&\text{ if }y^d=0^d,\\
[\a_k(1-\e)\ \e\ \overline{\a_k}\,\overline{\e}]&\text{ if }w_?(y^d)=k,
\end{cases}
\end{align}
where $w_?(y^d)$ denotes the number of erasures in $y^d$,
$\overline{x}\triangleq1-x$, and
$\overline{\a_k}=\overline{\a_0}/(1+k\overline{\a_0})$. Note that a
valid $y^d$ can have at most a single 1. Using Lagrangian techniques,
the choice of a common parameter $\a_k$ for all $y^d$ with
$w_?(y^d)=k$ can be shown to be optimal for this case, and
we skip the details. 

There are two cycles in $G^d_{d,\infty}$ - (1) the length-1 cycle
associated with $(u_00)$, and (2) the length-$(d+1)$ cycle associated
with $(u_0\,1\,u_0)$. Equating the length-normalized cycle metrics, we
get the constraint
\begin{equation}
  \frac{T(u_01)+\sum_{i=1}^dT(u_i0)}{d+1}=T(u_00).
\label{eq:19}
\end{equation}
Note that the metrics are functions of the parameter $\a_0$ and the
erasure probability $\e$.

By the dual bound corollary, we get the bound
\begin{equation}
  C_{(d,\infty)}(p_{Y|X})\le \min_{\substack{q:\\T(u_01)+\sum_{i=1}^dT(u_i0)=(d+1)T(u_00)}}T(u_00).
\label{eq:14}
\end{equation}
The above constrained optimization can be solved by standard
Lagrangian techniques (as shown in the Appendix) resulting in the
upper bound in Theorem \ref{thm:becd}. 

\subsection{Binary Symmetric Channel - Proof of Theorem \ref{thm:bsc}}
We will use the memory-1 Markov test distribution
$q(y_2|y_1)$ defined as 
\begin{equation}
   q(y_2|y_1)=\begin{bmatrix}
a&1-a\\
b&1-b
\end{bmatrix},
\label{eq:17}
\end{equation}
where the rows correspond to $y_1=0,1$, columns correspond to $y_2=0,1$
in that order, and $a,b$ are parameters satisfying
$0\le a,b\le1$. 

Using the dual bound corollary, we get the bound
\begin{align}
  C_{1,\infty}(p)\le\min_{a,b:T(01)+T(10)=2T(00)}T(00).\label{eq:8}
\end{align}
For the choice of test distribution over the binary symmetric channel BSC$(p)$, the metrics
evaluate to the following: 
\begin{align*}
  T(00)&=\op^2\log_2\frac{\op}{a}+p\op\log_2\frac{p\op}{b\ova}+p^2\log_2\frac{p}{\ovb},\\
  T(01)&=\op^2\log_2\frac{\op}{\ova}+p\op\log_2\frac{p\op}{a\ovb}+p^2\log_2\frac{p}{b},\\
  T(10)&=\op^2\log_2\frac{\op}{b}+p\op\log_2\frac{p\op}{a\ovb}+p^2\log_2\frac{p}{\ova},
\end{align*}
where $\overline{x}\triangleq 1-x$. Further, the constraint simplifies
as follows:
\begin{align}
&T(10)+T(01)-2T(00)=(1-2p)\left[(1-2p)\log_2\frac{a(1-b)}{b(1-a)}-\log_2\frac{1-b}{a}\right]=0.
\end{align}
Therefore, the optimization in \eqref{eq:8} simplifies to a 
problem in two variables $a,b$ taking values in $[0,1]$. This problem
is solved using standard Lagrangian methods to obtain the
bound in Theorem \ref{thm:bsc}.

\subsection{$(1,\infty)$-constrained binary-input AWGN Channel}
For $x\in\{0,1\}$, let $s(x)\triangleq(-1)^x$ denote the standard BPSK
modulation. Recall the notation for the Gaussian PDF
$\psi_{\mu,\sigma}=e^{-(x-\mu)^2/(2\sigma^2)}/(\sqrt{2\pi}\sigma)$ and
its integral $\Psi_{\mu,\sigma}(a,b)=\int_a^b\psi_{\mu,\sigma}(y)dy$ introduced earlier. 

As can be expected, the computations for the binary-input Gaussian channel,
defined by the conditional PDF $p_{Y|X}(y|x)=\psi_{s(x),\sigma}(y)$, are highly numerical in
nature. We use the dual bound corollary with the class of memory-1 Markov test
distribution defined by the conditional PDF $q(y_2|y_1)$ given in \eqref{eq:12} and depicted in
Fig. \ref{fig:qy}. 

The relative entropy
$D\bigl(p_{Y|X}(\,\cdot\,|x)\|q(\,\cdot\,|y_1)\bigr)$, denoted as $D_x(y_1)$, simplifies to
the following:
\begin{align}
D_x(y_1)/\log_2e=&\Psi_{s(x),\sigma}(-\infty,d_2(y_1,\sigma))\log_e\frac{e^{\frac{1+s(x)}{\sigma^2}}\Psi_{-1,\sigma}(-\infty,d_2(y_1,\sigma))}{a(y_1,\sigma)}+\nonumber\\
&\Psi_{s(x),\sigma}(d_2(y_1,\sigma),d_1(y_1,\sigma))\log_e\frac{\Delta(y_1,\sigma)}{\sqrt{2\pi}\sigma
  e^{\frac{s(x)}{2}}b(y_1,\sigma)}+\nonumber\\
&\Psi_{s(x),\sigma}(d_1(y_1,\sigma),\infty)\log_e\frac{e^{\frac{1-s(x)}{\sigma^2}}\Psi_{1,\sigma}(d_1(y_1,\sigma),\infty)}{c(y_1,\sigma)}+\nonumber\\
&\left(\frac{d_1(y_1,\sigma)-1}{2}-i(x)\right)\psi_{s(x),\sigma}(d_1(y_1,\sigma))-\left(\frac{d_2(y_1,\sigma)+1}{2}+\overline{i(x)}\right)\psi_{s(x),\sigma}(d_2(y_1,\sigma)),\nonumber
\end{align}
where $i(x)$ is $x$ seen as an integer. In the graph $G^1_{1,\infty}$, the metrics $T(x_1x_2)$, for $(x_1x_2)\in\{00,01,10\}$, are given by the following:
\begin{align}
  T(x_1x_2) &=\int_{-\infty}^{\infty}D_{x_2}(y_1)p_{Y|X}(y_1|x_1)dy_1\nonumber\\
&=\int_{-\infty}^{\infty}D_{x_2}(y_1)\psi_{s(x_1),\sigma}(y_1)dy_1.
\end{align}
The metrics above depend on the choice of the functions
$d_1$, $\Delta$, $a$ and $b$, which are all functions of $y_1$ and
$\sigma$. Note that $d_2=d_1-\Delta$ and $c=1-a-b$ are dependent on
the other functions. Also, the functions $a$, $b$ and $c$ take nonnegative
fractional values adding to 1, and the function $\Delta$ takes nonnegative values. 

The choice of the conditional PDF, its shape, its
piecewise nature, and its dependence on $y_1$ and $\sigma$ are
motivated by the $(1,\infty)$ constraint and the minimization of the
relative entropy in the dual upper bound. There are three piecewise
shapes for $q(y_2|y_1)$ - (1) shape of $N(-1,\sigma^2)$ for
$y_2<d_2(y_1,\sigma)$, (2) constant for $d_2(y_1,\sigma)\le y_2\le
d_1(y_1,\sigma)$, and (3) shape of $N(1,\sigma^2)$ for
$y_2>d_1(y_1,\sigma)$. These shapes are weighted by the fractions $a(y_1,\sigma)$,
$b(y_1,\sigma)$ and $c(y_1,\sigma)$.

If $y_1$ is large and
negative, then $x_1=1$ with high probability and this implies $x_2=0$
with high probability because of the $(1,\infty)$
constraint. Therefore, for large negative values of $y_1$,
$q(y_2|y_1)$ can be of the shape of $N(1,\sigma)$ for significant
values of $y_2$ by choosing the value of the function $d(y_1,\sigma)$
and the fractions $a(y_1,\sigma)$,
$b(y_1,\sigma)$ and $c(y_1,\sigma)$
appropriately. As the value of $y_1$ increases and becomes large and
positive, the probability of $x_1=0$ increases to 1. So, $x_2$ can be
0 or 1 with some nonzero probability as per the $(1,\infty)$
constraint. So, as $y_1$ increases, the function $d(y_1,\sigma)$
needs to increase to positive values. There are similar heuristics used
to motivate the other aspects of $q(y_2|y_1)$. 

Consider the class of functions
\begin{align}
\cF_{\a_{[1:4]}}(y_1,\sigma)=\frac{\a_1e^{y_1/\sigma^2}+\a_2e^{-y_1/\sigma^2}}{\a_3e^{y_1/\sigma^2}+\a_4e^{-y_1/\sigma^2}},
\end{align}
where $\a_i$, $1\le i\le 4$, are real-valued parameters. After some
experimentation, we have found that the 
functions $d_1$, $\Delta$, $a$ and $b$ can be chosen from the above class of functions with some suitable restrictions
on the values of parameters $\alpha_{[1:4]}$. The $\tanh$-like choice is
motivated by the form of the posterior probabilities over the BIAWGN
channel, which have the $\tanh$ form. In our computations, we use the following
set $\cF$ for the choice of the functions $d_1$, $\Delta$, $a$ and $b$:
\begin{align}
\cF=\{&d_1,\Delta,a,b:\nonumber\\
  &d_1\in \cF_{\delta_{[1:4]}}(y_1,\sigma),\,\,\delta_2\in[-1,1], \delta_1,\delta_2,\delta_3\in[0,1],\nonumber\\
  &\Delta\in \cF_{\Delta_{[1:4]}}(y_1,\sigma),\,\,\Delta_i\in[0,1],\nonumber\\
  &a\in \cF_{\a_{[1:4]}}(y_1,\sigma),\,\,\a_i\in[0,1],\nonumber\\
  &b\in \cF_{\b_{[1:4]}}(y_1,\sigma),\,\,\b_i\in[0,1],\nonumber\\
  &\max\left(\frac{\a_1}{\a_3},\frac{\a_2}{\a_4}\right)+\max\left(\frac{\b_1}{\b_3},\frac{\b_2}{\b_4}\right)\le 1\}.\nonumber
\end{align}
The above choices allow $d_1$ to be negative and ensures that $\Delta$
is positive and that $a$, $b$ and $c=1-a-b$ are nonnegative fractions
adding to 1. 

Using the dual bound corollary, we obtain the bound
\begin{equation}
  C_{(1,\infty)}(\sigma)\le \min_{\substack{d_1,\Delta,a,b\in\cF:\\T(01)+T(10)=2T(00)}}T(00).
\label{eq:15}
\end{equation}
The above problem is non-linear, and local minima can be found using
numerical optimization procedures. This bound is plotted in
Fig. \ref{fig:bpsk} as the $(1,\infty)$-constrained upper bound. The
lower bound is by using the method of \cite{1661831}. 

For the unconstrained BIAWGN$(\sigma^2)$ channel, the dual upper bound
is evaluated using the following test distribution
\begin{equation}
  q(y)=\begin{cases}
\dfrac{1-a(\sigma)}{2\Psi_{-1,\sigma}(-\infty,-\Delta(\sigma))}\,\psi_{-1,\sigma}(y),&y<-\Delta(\sigma),\\[10pt]
\dfrac{a(\sigma)}{2\Delta(\sigma)},&|y|\le\Delta(\sigma),\\[10pt]
\dfrac{1-a(\sigma)}{2\Psi_{1,\sigma}(\Delta(\sigma),\infty)}\,\psi_{1,\sigma}(y),&y>\Delta(\sigma).
\end{cases}
\end{equation}
parameterized by a positive real-valued function $\Delta(\sigma)$ and $a(\sigma)$ is a
function taking values in $[0,1]$ and will be chosen to get the best bound. The relative entropy
$D\bigl(p_{Y|X}(\,\cdot\,|x)\|q(\,\cdot\,)\bigr)$ for both $x=0,1$
simplifies to the following expression: 
\begin{align}
D\bigl(p_{Y|X}(\,\cdot\,|x)\|q(\,\cdot\,)\bigr)/\log_2e=&\bigl(1-\Psi_{1,\sigma}(-\Delta,\Delta)\bigr)\log_e\frac{2\Psi_{1,\sigma}(\Delta,\infty)}{1-a}+\Psi_{1,\sigma}\bigl(-\Delta,\Delta\bigr)\log_e\frac{2\Delta}{a\sqrt{2\pi}\sigma}+\label{eq:16}\\
&\qquad\frac{2}{\sigma^2}\Psi_{1,\sigma}(-\infty,-\Delta)+2\psi_{1,\sigma}(-\Delta) +\nonumber\\
&\frac{1}{2}\left[(\Delta-1)\psi_{1,\sigma}(\Delta)+(\Delta+1)\psi_{1,\sigma}(-\Delta) -\Psi_{1,\sigma}(-\Delta,\Delta)\right],\nonumber
\end{align}
where the dependence of $\Delta$ and $a$ on $\sigma$ is suppressed in
the notation to reduce clutter. From the above, it is seen that the
choice 
\begin{equation}
  a(\sigma)=\Psi_{1,\sigma}(-\Delta(\sigma),\Delta(\sigma))
\label{eq:20}
\end{equation}
minimizes the relative entropy. The upper bound on the capacity of the
unconstrained BIAWGN channel, shown in Fig. \ref{fig:bpsk}, is
obtained by setting $a(\sigma)$ as in \eqref{eq:20} and numerically finding the
best $\Delta(\sigma)$ that minimizes \eqref{eq:16}.
\section{Concluding remarks}
\label{sec:concluding-remarks}
The dual capacity bound is useful in scenarios where
characterizing the exact capacity is difficult. In particular,
restricting the test distributions to those that satisfy the
Karush-Kuhn-Tucker (KKT)
conditions on the capacity-achieving output distribution appears to be
an important idea for obtaining tight bounds with simple
characterizations. In this paper, KKT-constrained test distributions
were explored for runlength constrained binary channels.

For runlength constrained channels, the
KKT constraint is converted into a condition on the metrics of cycles
in the state diagram. For larger memory of the test distribution, the state diagram has many
cycles and computation complexity of the method increases, but,
interestingly, the bounds for low memory appear to be tight in many
cases of theoretical and practical interest. Characterizing the gap
between the upper bound and achievable rates 
analytically is an interesting problem to pursue in the
future. Extending the method to other channels with memory, such as
the Inter Symbol Interference (ISI) channel, is another interesting
avenue for future work.  

\bibliographystyle{IEEEtran}
\bibliography{refs,references}

\begin{thebibliography}{10}
\providecommand{\url}[1]{#1}
\csname url@samestyle\endcsname
\providecommand{\newblock}{\relax}
\providecommand{\bibinfo}[2]{#2}
\providecommand{\BIBentrySTDinterwordspacing}{\spaceskip=0pt\relax}
\providecommand{\BIBentryALTinterwordstretchfactor}{4}
\providecommand{\BIBentryALTinterwordspacing}{\spaceskip=\fontdimen2\font plus
\BIBentryALTinterwordstretchfactor\fontdimen3\font minus
  \fontdimen4\font\relax}
\providecommand{\BIBforeignlanguage}[2]{{%
\expandafter\ifx\csname l@#1\endcsname\relax
\typeout{** WARNING: IEEEtran.bst: No hyphenation pattern has been}%
\typeout{** loaded for the language `#1'. Using the pattern for}%
\typeout{** the default language instead.}%
\else
\language=\csname l@#1\endcsname
\fi
#2}}
\providecommand{\BIBdecl}{\relax}
\BIBdecl

\bibitem{handbook}
B.~H. Marcus, R.~M. Roth, and P.~H. Siegel, ``Constrained systems and coding
  for recording channels,'' in \emph{Handbook of Coding Theory}, V.~S. Pless
  and W.~C. Huffman, Eds.\hskip 1em plus 0.5em minus 0.4em\relax Amsterdam:
  Elsevier, 1998, pp. 1635--1764.

\bibitem{1661831}
D.~M. Arnold, H.~A. Loeliger, P.~O. Vontobel, A.~Kavcic, and W.~Zeng,
  ``Simulation-based computation of information rates for channels with
  memory,'' \emph{IEEE Transactions on Information Theory}, vol.~52, no.~8, pp.
  3498--3508, Aug 2006.

\bibitem{955166}
P.~Vontobel and D.~Arnold, ``An upper bound on the capacity of channels with
  memory and constraint input,'' in \emph{Information Theory Workshop, 2001.
  Proceedings. 2001 IEEE}, 2001, pp. 147--149.

\bibitem{6875399}
Y.~Li and G.~Han, ``Input-constrained erasure channels: Mutual information and
  capacity,'' in \emph{Information Theory (ISIT), 2014 IEEE International
  Symposium on}, June 2014, pp. 3072--3076.

\bibitem{7308065}
O.~Sabag, H.~H. Permuter, and N.~Kashyap, ``The feedback capacity of the binary
  erasure channel with a no-consecutive-ones input constraint,'' \emph{IEEE
  Transactions on Information Theory}, vol.~62, no.~1, pp. 8--22, Jan 2016.

\bibitem{topsoe67}
F.~Tops{\o}e, ``An information theoretical identity and a problem involving
  capacity,'' \emph{Studia Sci. Math. Hungar.}, vol.~2, pp. 291--292, 1967.

\bibitem{KEMPERMAN1974101}
J.~Kemperman, ``On the {S}hannon capacity of an arbitrary channel,''
  \emph{Indagationes Mathematicae (Proceedings)}, vol.~77, no.~2, pp. 101 --
  115, 1974.

\bibitem{Csiszar11}
{I. Csisz\'ar and J. K\"orner}, \emph{Information Theory: Coding Theorems for
  Discrete Memoryless Systems}.\hskip 1em plus 0.5em minus 0.4em\relax
  Cambridge University Press, 2011.

\bibitem{LMW2009}
A.~Lapidoth, S.~Moser, and M.~Wigger, ``On the capacity of free-space optical
  intensity channels,'' \emph{IEEE Trans. Inf. Theory}, vol.~55, no.~10, pp.
  4449--4461, Oct 2009.

\bibitem{1237131}
A.~Lapidoth and S.~M. Moser, ``Capacity bounds via duality with applications to
  multiple-antenna systems on flat-fading channels,'' \emph{IEEE Transactions
  on Information Theory}, vol.~49, no.~10, pp. 2426--2467, Oct 2003.

\bibitem{6218668}
G.~Durisi, ``On the capacity of the block-memoryless phase-noise channel,''
  \emph{IEEE Communications Letters}, vol.~16, no.~8, pp. 1157--1160, August
  2012.

\bibitem{7282870}
A.~Thangaraj, G.~Kramer, and G.~Böcherer, ``Capacity upper bounds for
  discrete-time amplitude-constrained awgn channels,'' in \emph{2015 IEEE
  International Symposium on Information Theory (ISIT)}, June 2015, pp.
  2321--2325.

\bibitem{bang2013digraphs}
J.~Bang-Jensen and G.~Gutin, \emph{Digraphs: Theory, Algorithms and
  Applications}.\hskip 1em plus 0.5em minus 0.4em\relax Springer London, 2013.

\end{thebibliography}

\section*{Appendix: Langrangian computations}
\subsection{$(1,\infty)$-BEC$(\e)$: Theorem \ref{thm:bec} - Part (1)}
For the optimization problem in \eqref{eq:22}, the Lagrangian is
defined as follows:
\begin{align}
L&=T(00)+\lambda(T(01)+T(10)-2T(00)),\nonumber\\
&=(1-2\lambda)T(00)+\lambda T(01)+\lambda T(10).
\end{align}
The partial derivative of $L$ with respect to $\a$ simplifies as follows:
\begin{align}
  (\log_e2)\frac{\pd L}{\pd\a}&=(1-2\lambda)\frac{-\e(1-\e)}{\a}+\lambda\frac{\e(1-\e)}{\oa}+\lambda\frac{-\e(1-\e)}{\a}\nonumber\\
&=\e(1-\e)\left[\frac{\lambda}{1-\a}-\frac{1-\lambda}{\a}\right].
\end{align}
Equating to zero, we get
\begin{equation}
  \alpha=1-\lambda.
\label{eq:31}
\end{equation}
Similarly, equating the partial derivative of $L$ with respect to $\b$
to zero, we get
\begin{equation}
  \lambda=\frac{1-\b}{2-\b}.
\label{eq:32}
\end{equation}
Using \eqref{eq:31} and \eqref{eq:32} in the objective function
$T(00)$ and the constraint $T(00)+T(01)-2T(00)=0$, we get the
statement of Theorem \ref{thm:bec} - Part (1).

\subsection{$(1,\infty)$-BEC$(\e)$: Theorem \ref{thm:bec} - Part (2)}
The metrics in $G^2_{1,\infty}$ are as follows:
\begin{align}
  T(000)&=(1-\e)^3\log_2\frac{1}{\a_{00}}+\e(1-\e)^2\log_2\frac{1}{\a_{0?}}+\e(1-\e)^2\log\frac{1}{\a_{?0}}+\e^2(1-\e)\log_2\frac{1}{\a_{??}},\nonumber\\
  T(100)&=(1-\e)^3\log_2\frac{1}{\a_{10}}+\e(1-\e)^2\log_2\frac{1}{\a_{1?}}+\e(1-\e)^2\log\frac{1}{\a_{?0}}+\e^2(1-\e)\log_2\frac{1}{\a_{??}},\nonumber\\
  T(010)&=\e(1-\e)^2\log_2\frac{1}{\a_{0?}}+\e^2(1-\e)\log_2\frac{1}{\a_{??}},\nonumber\\
  T(001)&=(1-\e)^3\log_2\frac{1}{\oa_{00}}+\e(1-\e)^2\log_2\frac{1}{\oa_{0?}}+\e(1-\e)^2\log\frac{1}{\oa_{?0}}+\e^2(1-\e)\log_2\frac{1}{\oa_{??}},\nonumber\\
  T(100)&=(1-\e)^3\log_2\frac{1}{\oa_{10}}+\e(1-\e)^2\log_2\frac{1}{\oa_{1?}}+\e(1-\e)^2\log\frac{1}{\oa_{?0}}+\e^2(1-\e)\log_2\frac{1}{\oa_{??}}. \nonumber
\end{align}
The Lagrangian for the optimization problem in \eqref{eq:25} is
\begin{align}
  L=&(1-2\lambda_1-3\lambda_3)T(000)+(\lambda_1+\lambda_2)T(010)+\lambda_1T(101)+\lambda_2T(100)+\lambda_2T(001).
\end{align}
Equating the partial derivatives with respect to $\a_{rs}$,
$(rs)\in\{00,0?,?0,??,10,1?\}$, we obtain the following relationships:
\begin{align*}
  \a_{00}=1-\frac{\lambda_2}{1-2\l_1-2\l_2},\,\,\, &\a_{0?}=1-\frac{\l_2}{1-\l_1-\l_2},\\
\a_{?0}=1-\frac{\l_1+\l_2}{1-\l_1-\l_2},\,\,\, &\a_{??}=1-\l_1-\l_2,\\
\a_{10}=\frac{\l_2}{\l_1+\l_2},\,\,\, &\a_{1?}=\frac{\l_1}{\l_1+\l_2}.
\end{align*}
Setting $\b=\a_{00}$ and $\a=\a_{??}$, we express all variables in
terms of $\a$ and $\b$. Expressing the objective function 
and the constraints in terms of $\a$ and $\b$ and simplifying results in the statement
of Theorem \ref{thm:bec} - Part (2).

\subsection{$(1,2)$-BEC$(\e)$: Theorem \ref{thm:bec12}}
\subsubsection{Part (1)}
The metrics in $G^1_{1,2}$ are as follows:
\begin{align}
T(001)&=\e(1-\e)^2\log_2\frac{1}{\oa_{0?}}+\e(1-\e)^2\log_2\frac{1}{\oa_{?0}}+\e^2(1-\e)\log_2\frac{1}{\oa_{??}}\nonumber\\
T(010)&=\e(1-\e)^2\log_2\frac{1}{\a_{0?}}+\e^2(1-\e)\log_2\frac{1}{\a_{??}}\nonumber\\
T(100)&=(1-\e)^3\log_2\frac{1}{\a_{10}}+\e(1-\e)^2\log_2\frac{1}{\a_{?0}}+\e(1-\e)^2\log_2\frac{1}{\a_{1?}}+\e^2(1-\e)\log_2\frac{1}{\a_{??}}\nonumber\\
T(101)&=(1-\e)^3\log_2\frac{1}{\oa_{10}}+\e(1-\e)^2\log_2\frac{1}{\oa_{?0}}+\e(1-\e)^2\log_2\frac{1}{\oa_{1?}}+\e^2(1-\e)\log_2\frac{1}{\oa_{??}}\nonumber
\end{align}
The Lagrangian for the optimization problem in \eqref{eq:29} can be
written as
\begin{align}
  L&=\frac{1}{2}(T(010)+T(101))+\l(T(010)+3T(101)-2T(100)-2T(001))\nonumber\\
&=(\frac{1}{2}+\l)T(010)+(\frac{1}{2}+3\l)T(101)-2\l T(100)-2\l T(001).
\end{align}
Equating partial derivatives of $L$ with respect to the parameters
$\a_{rs}$, $(rs)\in\{10,0?,?0,1?,??\}$, we get
\begin{align*}
  \a_{10}=\a_{1?}=\frac{4\l}{1-2\l},\,\, &\a_{0?}=\frac{1-2\l}{1+2\l},\\
\a_{?0}=\frac{4\l}{1+2\l},\,\,&\a_{??}=\frac{1}{2}+\l.
\end{align*}
Setting $\b=\oa_{10}=\frac{1-6\l}{1-2\l}$, we express all variables in
terms of $\b$. Expressing the objective function and the constraint in
terms of $\b$ and simplifying results in the statement of Theorem \ref{thm:bec12} -
Part (1). 

\subsubsection{Part (2)}
The metrics in $G^2_{1,2}$ are as follows:
\begin{align}
 &T(0101)=(1-\e)\left[\e^3 \log_2\frac{1}{\oa_{???}}+\e^2(1-\e)(\log_2\frac{1}{\oa_{0??}}+\log_2\frac{1}{\oa_{??0}})+\right.\nonumber\\
&\qquad\qquad\qquad \qquad\qquad\qquad\left.((1-\e)^3+3\e(1-\e)^2+\e^2(1-\e))\log_2\frac{1}{\oa_{010}}\right],\nonumber\\
 &T(0100)=(1-\e)\left[\e^3 \log_2\frac{1}{\a_{???}}+\e^2(1-\e)(\log_2\frac{1}{\a_{0??}}+\log_2\frac{1}{\a_{??0}})+\right.\nonumber\\
&\qquad\qquad\qquad \qquad\qquad\qquad\left.((1-\e)^3+3\e(1-\e)^2+\e^2(1-\e))\log_2\frac{1}{\a_{010}}\right],\nonumber\\
&T(1010)=(1-\e)\left[\e^3 \log_2\frac{1}{\a_{???}}+\e(1-\e)^2\log_2\frac{1}{\a_{10?}}+\e^2(1-\e)(\log_2\frac{1}{\a_{?0?}}+\log_2\frac{1}{\a_{1??}})\right],\nonumber\\
&T(0010)=(1-\e)\left[\e^3 \log_2\frac{1}{\a_{???}}+\e^2(1-\e)(\log_2\frac{1}{\a_{0??}}+\log_2\frac{1}{\a_{?0?}})\right],\nonumber\\
&T(1001)=(1-\e)\left[\e^3 \log_2\frac{1}{\oa_{???}}+\e(1-\e)^2\log_2\frac{1}{\oa_{10?}}+\e^2(1-\e)(\log_2\frac{1}{\oa_{??0}}+\log_2\frac{1}{\a_{?0?}}+\log_2\frac{1}{\a_{1??}})\right]. \nonumber
\end{align}
The Lagrangian for the optimization problem in \eqref{eq:30} can be
written as follows:
\begin{align}
 L&=\frac{1}{2}(T(0101)+T(1010))+\l(2T(0010)+2T(0100)+2T(1001)-3T(0101)-3T(1010))\nonumber\\
&=(\frac{1}{2}-3\l)(T(0101)+T(1010))+2\l(T(0010)+T(0100)+T(1001)).
\end{align}
Equating partial derivatives of $L$ with respect to the parameters
$\a_{rst}$, $rst\in\{010,10?,0??,?0?,??0,1??,???\}$, we get
\begin{align*}
  \a_{010}=\frac{4\l}{1-2\l},\,\,&\a_{10?}=\a_{1??}=\frac{1-6\l}{1-2\l},\\
\a_{0??}=\frac{8\l}{1+2\l},\,\,&\a_{??0}=\frac{4\l}{1+2\l},\\
\a_{?0?}=\frac{1-2\l}{1+2\l},\,\,&\a_{???}=\frac{1}{2}+\l.
\end{align*}
Setting $\b=\oa_{010}=\frac{1-6\l}{1-2\l}$, we express all variables in
terms of $\b$. Expressing the objective function and the constraint in
terms of $\b$ and simplifying results in the statement of Theorem \ref{thm:bec12} -
Part (2). 

\subsection{$(d,\infty)$-BEC$(\e)$: Theorem \ref{thm:becd}}
The metrics in the graph $G^d_{d,\infty}$ can be written as follows:
\begin{align}
  T(u_00)&=\sum_{k=0}^d\binom{d}{k}\e^k(1-\e)^{d-k+1}\log_2\frac{1}{\a_k},\nonumber\\
  T(u_i0)&=\sum_{k=1}^d\binom{d-1}{k-1}\e^k(1-\e)^{d-k+1}\log_2\frac{1}{\a_k},\nonumber\\
  T(u_01)&=\sum_{k=0}^d\binom{d}{k}\e^k(1-\e)^{d-k+1}\log_2\frac{1}{\oa_k},\nonumber
\end{align}
The Lagrangian for the optimization problem in \eqref{eq:14} is
\begin{align}
  L=T(u_00)+\l(T(u_01)+\sum_{i=1}^dT(u_i0)-(d+1)T(u_00)).
\end{align}
Equating partial derivatives of $L$ with respect to the parameters
$\a_k$, $0\le k\le d$, we get
\begin{align}
  \a_k=\frac{(d-k+1)\l-1}{(d-k)\l-1}.
\end{align}
Expressing all variables in terms of $\a_0$ and simplifying the
objective function and the constraint results in the statement of
Theorem \ref{thm:becd}.
\section*{Acknowledgements}
The author thanks Navin Kashyap and Gerhard Kramer for useful
discussions on the topic of this article.

\end{document}